

\input tables.tex
\magnification=1200
\hsize=6.truein
\vsize=8.5truein
\settabs 18\columns
\baselineskip=14pt
\def\b{\bigskip}
\def\bb{\bigskip\bigskip}

\def\s{\smallskip}
\def\th{\thinspace}

\def\no{\noindent}
\def\ce{\centerline}
\def\rt{\rightline}

\def\leaderfill{\ \leaders\hrule\hfill\ }
\def\x{${\tilde{\chi}}^0_i$}
\font\stephalf=cmr8 scaled\magstephalf

\def\date{\no{\ifcase\month
        \or January \or February\or March\or April\or May\or June\or
        July\or August\or September\or October\or November\or December\fi
        \space\number\day, \number\year}\bigskip\bigskip\bigskip}

\def\draftprep #1\par{\no\underbar{\tt DRAFT}\par\rt{$\scriptstyle\rm{#1}$}
        \vskip0.25truecm\par}
\def\prep #1\par{\rt{$\scriptstyle\rm{#1}$}\vskip1truecm\par}
\def\ttl #1 \par{\ce{\bf\uppercase{#1}}\vskip1.5truecm}
\def\ttls #1 {\ce{\bf\uppercase{#1}}}\par
\def\fttl #1 #2 {\ce{\bf\uppercase{#1}\footnote\dag{#2}}\par\b}
\def\gg{\b{\ce{\bf GRACIELA GELMINI}}\s}
\def\phys {\baselineskip=12pt{\hsize=16truecm
   \ce{\it Department of Physics, University of California at Los Angeles}
   \ce{\it 405 Hilgard Avenue, Los Angeles, California 90024-1547}
   \s }}

\def\abttl {\vskip0.75truecm{\ce{\bf ABSTRACT}}\s}

\def\ab #1 {{\baselineskip=10pt{\ce{\vbox{\hsize=5truein
 { \stephalf{#1} }}}}\s}}

\def\sects #1 #2 {\b\ce{\bf
                        \vtop{\hsize=0.75truecm{\no #1\hfill}}\hfill
                        \vtop{\hsize=15truecm{\no{#2}}}}\par\b}

\def\ssects #1 #2 {\s\ce{\it
                        \vtop{\hsize=1truecm{\no {\it #1}\hfill}}\hfill
                        \vtop{\hsize=14.90truecm{\no{\it{#2}}}}}\par}

\def\sect #1 #2 {\b\ce{{#1}\ \uppercase{#2}}\par\b}

\def\fs #1 {$^{\bf{#1}}$}

\def\conts #1 #2 #3\par{\line{\vtop{\hsize=1truecm
                        \no{#1}).}\hfill
                        \vtop{\hsize=15.25truecm\no{{#2}
                        \th\leaderfill\th {#3}}
                        }}\par}\s

\def\rf #1 #2 {{\baselineskip=10pt\line{
                    \vtop{\hsize=0.75truecm\noindent{{#1}.}}\hfill
                    \vtop{\hsize=15truecm{\no{#2}}} }\par}}
\def\apndxttl #1 {\vskip2truecm{\ce{\underbar{APPENDIX} {#1}}}\par\bb}

\def\cont #1 #2 #3{\par{\baselineskip=10pt{\line{\vtop{\hsize=1truecm
                        \no{#1}}\hfill
                        \vtop{\hsize=14.75truecm\no{{#2}
                        \th\leaderfill\th {#3}}
                        }}}\par}\s}

\ttl{Beyond \ the \ Standard \ Model}

\gg

\phys

\abttl

\ab{ A pedagogical introduction is given to attempts to formulate a
more fundamental theory of elementary particles than that provided
by the standard model. After a review of the basic elements of the
 standard model, its unsatisfactory features, and ideas to modify
 them will be discussed. An introduction to Technicolour, t-quark
  condensates, supersymmetric models, GUT's and a brief comment
on composite models follow.}
\b
\no {\bf 1. ~Introduction}
\s
The success of the standard model (S.M.) as the theory of elementary
particles is indisputable.  This model provides the answers to many of
the fundamental questions formulated in the `60's:

\item{-} Weak interactions are renormalizable when unified with
electromagnetism in the Glashow-Weinberg-Salam theory$^1$.
\s
\item{-}  Quarks have a physical reality, as elementary particles
confined into hadrons.
\s
\item{-} A Lagrangian theory of strong interactions can be formulated;
this is QCD$^2$.
\s
\item{-} There is a unity in the description of particle interactions
(except gravity), since they can be described by gauge theories, with the
profound distinction among them depending on whether the vacuum state
respects the symmetry or not.
\s
All experimental and theoretical tests until now verify that these ideas
are correct.  At this point discrepancies would be more interesting than
confirmations. Thus, why look for physics beyond the S.M.?
The main reason is that by thinking about what may come next,
we improve our chances of finding it.  We may design the experimental
and theoretical tests that uncover it, and  we may learn how to
interpret possible signatures of new physics.  Another reason is that
the S.M. has unsatisfactory features and leaves open questions that may
be solved by further use of some of the successful ideas of the S.M., or
by entirely new ideas.  The main new ideas incorporated into the S.M.
are:
\s
\item{-} {\underbar{A new layer of compositeness}: ``preons" may constitute
quarks, in the same way quarks constitute hadrons. Preons may even yield
dynamical symmetry breaking of the electroweak symmetry (Technicolour models),
repeating at higher energies the breaking of chiral symmetry at the color
confinement scale.
\s
\item{-}{\underbar{More symmetry}: Grand Unified symmetries may repeat at
higher energy scales the electroweak unification, this time unifying also
strong interactions, or we may find, instead, a boson-fermion symmetry, i.e. a
supersymmetry.
\s
\no An example of entirely new ideas not contained in the S.M., is provided by
string theories. String models are based on a radical
modification of the concepts on which the S.M. rests as a local quantum field
theory, since the fundamental degrees of freedom are not located at points in
space-time but in one dimensional extended objects, the strings. However,
effective local field theories are derived from string theories,  as their low
energy limit. These theories are extremely interesting. Superstring theories
are the only candidates at present for a unified theory of all interactions
including gravity. Unfortunately, at present, an  infinite variety of  gauge
models and matter spectra seem to be consistent with perturbative solutions to
string theories. We are very far from formulating and solving a unique,
predictive string theory.

\b
The definition of physics ``beyond" the S.M. rests on what we mean by
the S.M. itself.  We will start these lectures, therefore, with an
overview of the main features of the S.M. including a review
of gauge and global interactions, exact and spontaneously broken, and
the possible breaking mechanisms.  We will then mention the
unsatisfactory features of the S.M. and possible ways to
modify them.  We will review current attempts to incorporate
either more symmetry or a new layer of compositeness.  We will talk
about the possibility of a dynamical symmetry breaking of the
electroweak symmetry, both in Technicolour models (T.C.) and through
top quark condensates.  We will present the minimal supersymmetric
extension of the standard model (M.S.S.M.) and mention minimally
non-minimal extensions.  We will talk about grand unified theories
(GUT's) and supersymmetric GUT's and we will briefly mention composite
models.

These lectures follow a pedagogical approach.  I will not attempt an
exhaustive comparison with data.
\b
\no {\bf 2.~The Standard Model}
\b
The S.M. is a gauge theory.  Any gauge theory has three defining
elements:
\s
\no i) {\underbar{The symmetry group.}}  It is $SU_{\rm Colour}(3)
\times SU_{\rm Left} (2) \times U_Y (1)$, where $Y$ is the weak
hypercharge.
\s
\no ii) {\underbar{The representation of the matter fields,}} the
fermions $f$.  There are 15 Weyl \break spinors,  repeated twice at
larger masses, yielding three generations or families (see Table 1.).
In Table 1., the parentheses show the representation of $SU_{C}(3)$, the
representation of $SU_L(2)$ (both indicated by their dimensionality), and the
value of $Y$, for each set of fields.  Here, the index $i = 1,2,3$ is the
colour index. The sub-indices $L$ and $R$ indicate Weyl spinors of
left and right-handed chirality (see
later).  Notice that the right-handed neutrinos
$\nu_{e_{R}}, \nu_{\mu_{R}}, \nu_{\tau_{R}}$ are
absent in the S.M. (in order to prevent the appearance of Dirac neutrino
masses).  They could easily be added as singlets under the three groups,
$\nu_R :~(1,1,0)$.
\bb
\no Table 1. Matter fields in the Standard Model.
\s

\vbox{\offinterlineskip
\hrule
\halign{&\vrule#&
\strut\quad\hfil#\quad\cr
height2pt&\omit&&\omit&\cr
&{\bf Quarks}\hfil&&{\bf Leptons}\hfil&\cr
height2pt&\omit&&\omit&\cr
\noalign{\hrule}
height2pt&\omit&&\omit&\cr
&$q_L :~ (3,2,1/6)$\hfil&& $\ell_L :~(1,2, -1/2)$\hfil &\cr
&&&& \cr
&$\pmatrix{u \cr d\cr}_{Li}~,~~\pmatrix{c \cr s\cr}_{Li}~,~~
\pmatrix{t \cr b\cr}_{Li}$&&$\pmatrix{\nu_e\cr e\cr}_L~,~~
\pmatrix{\nu_{\mu} \cr \mu\cr}_L~,~~ \pmatrix{\nu_{\tau}\cr
\tau\cr}_L$&\cr
&&&&\cr
& $u_R :~ (3,2,2/3)$\hfil~~~&& &\cr
& $u_{Ri}~,~~ c_{Ri}~,~~ t_{Ri}$ \hfil~~~&& &\cr
&&&& \cr
& $d_R :~ (3, 1, -1/3)$\hfil && $e_R :~ (1, 1, -1)$ \hfil & \cr
& $d_{R_i}~,~~ s_{Ri}~,~~ b_{Ri}$ \hfil && $e_R~,~~ \mu_R~ ,~~ \tau_R$ \hfil
& \cr height2pt&\omit&&\omit&\cr}
\hrule}
\b
\no
\s
\no The fields $f_L~ (q_L$ and $\ell_L$) have weak isospin $T_L = 1/2$, thus
its third component $T_{3L} $ is $+1/2$ and $-1/2$ for the upper and lower
components respectively.  The other fields, the $f_R$, have $T_L= 0$, thus
$T_{3L} = 0$.  The charge $Q$ of each field is given by
$$ Q = T_{3L} + Y ~. \eqno{(1)}$$
The charges of the particles in each generation sum to zero, $\sum_f Q = 0$.
This
insures that anomalies cancel within each family and, thus, the S.M. symmetry
is preserved at the quantum level.  The $SU_L(2) \times U_Y(1)$ assignments of
$f_L$ and $f_R$ make it impossible to have mass terms
$$ m_f ({\bar{f}}_L ) f_R \eqno{(2)}$$
in the Lagrangian (since only terms that are singlet, i.e. invariant under the
symmetry group, can appear in the Lagrangian).
\s
\no  iii)~{\underbar{A mechanism for spontaneous symmetry breaking}.
In the S.M. there is a fundamental scalar  particle $\phi : (1,2, -
1/2)$ with a potential $V (\phi^+ \phi)$ that has a minimum at
$$ < \phi >  \not= 0~. \eqno(3)$$
The non-zero vacuum expectation value (V.E.V.)  $ < \phi > $
breaks the electroweak symmetry into electromagnetism
$$ SU_L (2) \times U_Y (1) \to U_{em} (1) \eqno(4)$$
and induces terms as in Eq. 2, that give masses to the quarks and
leptons, except neutrinos.
\b
Let us analyze the main features of the S.M. contained in a compact way
in the three points above.
\b
\no  {\it 2.1. ~Weyl Spinors}
\s
The electroweak interactions
distinguish left-handed from right-handed  \break fermions, where the
chirality
components $L$ and $R$ of a Dirac fermion $f$ are defined as
$$ f_L = P_L f = {1 - \gamma_5\over 2}  f~,~~f_R = P_R f = { 1 +
\gamma_5\over 2} f~. \eqno(5)$$
As we have seen above, in the S.M. the fields $f_L$ are doublets
and the fields
$f_R$ are singlets of $SU_L (2)$.  This different assignment can
 be done
because gauge interactions conserve chirality, and it is done to
 account for
maximal parity violation in weak interaction.

A Dirac fermion $f$ has four independent complex components;  $f_L$ and
$f_R$ are two orthogonal pieces of $f$, each a two-component Weyl
spinor,

$$ f = {1 + \gamma_5 + 1 - \gamma_5\over 2} = P_L f + P_R f = f_L +
f_R~.
\eqno(6)$$
We will frequently use the following properties of the
$\gamma$-matrices$^3$:
$$ \gamma_5^2 = 1~,~~ \gamma^+_5 = \gamma_5~,~~ \gamma_5 \gamma_\mu = -
\gamma_{\mu} \gamma_5~. \eqno(7)$$
We can, thus, prove that $P_L$ and $P_R$ are projectors (i.e. $P^2_L
= P_L,~
P^2_R = P_R)$ onto orthogonal components (i.e. $P_L P_R = P_R P_L = 0)$.
  For example,
$$\eqalignno{
P^2_L =& {{\left( 1-\gamma_5\right)}\over{4}}^2 = {{1\th -\th 2\gamma_5\th
+\th
\gamma^2_5}\over{4}}={{1\th -\th\gamma_5}\over{2}}= P_L~, & (8)\cr
P_L\th P_R =& {{\left( 1\th -\th\gamma_5\right)}\over{2}}\th {{\left( 1\th
+\th\gamma_5\right)}\over{2}}={{1\th -\th\gamma^2_5}\over{4}}=0~. & (9)\cr}$$
\no On the conjugate fermions $\bar{f} \equiv f^{\dag}\th \gamma_0$, $P_L$
 and $P_R$ act in the following way:
$$\eqalignno{
\bar{f}\th P_L =& f^{\dag}\th \gamma_0P_L = f^{\dag}\th P_R\gamma_0 =
f^{\dag}\th P^{\dag}_R\th
\gamma_0 = \left( P_Rf\right)^{\dag}\th \gamma_0 =
\overline{\left( f_R\right)}~, &(10)\cr
\bar{f}\th P_R =& \overline{\left( f_L\right)}~. & (11)\cr}$$
\no We can see now that kinetic terms, ${\cal L}_k = i\bar{f}\gamma^\mu\th
\partial_\mu\th f$, and gauge interaction terms, \break ${\cal L}_{g.i.}= g\bar
f\gamma^\mu\th fA_{\mu}$, do not change chirality, due to the presence of
$\gamma^\mu$,
$$\bar{f}\gamma^{\mu}\th f = \overline{\left( f_L+f_R\right)}\th \gamma^\mu\th
\left( f_L\th +\th f_R\right) = \overline{\left( f_L\right)}\gamma_{\mu}
f_L\th +\th
\overline{\left( f_R\right)}\gamma_{\mu}\th f_R\eqno{(12)}$$
since the mixed terms vanish: $\overline{\left(
f_R\right)}\gamma_{\mu} f_L =
\bar{f}\th P_L\gamma^\mu\th P_R f = \bar{f}\gamma^\mu\th\left( P_R\th
P_L\right) f=0$.

Mass terms, instead, mix chirality. Dirac masses mix two different Weyl
spinors,
$$m^D\th\bar{f}f = m^D\overline{\left( f_L + f_R\right)}\th\left( f_L +
f_R\right) = m^D\left({\overline{\left( f_L\right)}f_R\th +\th
\overline{\left(
f_R\right)}f_L}\right)~.\eqno{(13)}$$
\no Majorana masses mix a Weyl spinor with itself,
$$m^M\th (f_L)^T\th Cf_L\equiv m^M\th \overline{\left( f^C\right)_R}
f_L~.
\eqno{(14)}$$
Since here a field is mixed with its charge-conjugate, this term is only
possible for particles that have zero charge of any additive conserved quantum
number, such as electric charge. Let us clarify Eq. (14), and
introduce concepts that
will be useful later, by examining the
charge conjugation $C$. The matrix $C$ is
defined as
$C=-C^{-1}\th =-C^{\dag}=-C^T=i\gamma^2\gamma^0$. Under charge conjugation
Dirac and Weyl fermions transform in the following way (see, for example Ref.
4):
$$\eqalignno{
f &\buildrel{C}\over{\longrightarrow} f^C \equiv C\bar f^T & (15)\cr
\bar f &\buildrel{C}\over{\longrightarrow} \overline{f^C} \equiv f^T\th C &
(16)
\cr
f_L &= P_Lf\buildrel{C}\over{\longrightarrow} P_L f^C = \left( f^C\right)_L =
C\overline{\left( P_R f\right)^T} = C\overline{\left( f_R\right)^T} & (17)\cr
f_R &-\hskip-.2truecm\buildrel{C}\over{-\hskip-.2truecm\longrightarrow}
\left( f^C\right)_R = C\overline{\left(
f_L\right)^T} & (18) \cr
\overline{\left( f_L\right)} &\buildrel{C}\over
{-\hskip-.2truecm\longrightarrow} \overline{\left( f^C\right)_L}
=(f_R)^T\th C &
(19)\cr
\overline{\left( f_R\right)}
&\buildrel{C}\over{-\hskip-.2truecm\longrightarrow} \overline{\left(
f^C\right)_R} = =(f_L)^T\th C & (20)\cr}$$
Notice that only two of the four Weyl spinors $f_L, f_R, \left(f^C\right)_L$
and $\th \left( f^C\right)_R$, are independent. Usually we choose $f_L,\th f_R$
as independent (then $\left( f^C\right)_R =$ $C\overline{\left(
f_L\right)^T}$, $ \left( f^C\right)_L =$ $C\overline{\left(
f_R\right)^T}~$). But in GUT's it is usually better to take both
independent Weyl spinors with the same chirality. For example $f_L$, $\left(
f^C\right)_L~ $ (then $\left( f^C\right)_R = $ $C \overline{\left(
f_L\right)^T}$, $f_R=C\overline{\left( f^C\right)^T_L}~$).
\b
\no {\it 2.2. ~Gauge Interactions}
\s
Starting from a Lagrangian invariant under a group G of continuous
\underbar{global} transformations, we obtain a Lagrangian invariant under
\underbar{local} (or gauge) transformations by replacing the derivatives
$\partial_\mu$ by covariant derivatives $D_\mu$. A Lagrangian, function of
fields $\chi_\alpha$ and their derivatives $\partial_\mu\chi_\alpha$ ($\alpha$
labels all fields),
$${\cal{L}} = {\cal{L}}\th \left(\chi_\alpha,\
\partial_\mu\chi_\alpha\right)~,\eqno {(21)}$$
\no is invariant under a global group G of transformations $U$, if
${\cal{L}}$, written
in terms of transformed fields
$$\chi^\prime_\alpha = U\chi_\alpha\th ,\qquad \partial_\mu\chi^\prime_\alpha =
U\partial_\mu\chi_\alpha~,\eqno{(22)}$$
\no is the same function in Eq. (21) of the new fields,
${\cal{L}}={\cal{L}}\left(
\chi^\prime_\mu,\th \partial_\mu\chi^\prime_\mu\right)$.

We obtain a gauge invariant Lagrangian by replacing
$$\partial_\mu \to D_\mu\ = \ \partial_\mu\th -\th ig\th \sum_a
\th A^a_\mu\th T^a~, \eqno{(23)}$$
where $A^a_\mu$ are new fields, the gauge fields, and $T^a$ are the
generators of the group G. There are as many gauge fields as generators (both
labelled by the index $a$). Any transformation $U$ can be written in terms of
the generators $T^a$ as $U = {\rm exp} \{ i \sum_a\th\alpha^a T^a \}$ where
 $\alpha^a$
are free parameters.  In \underbar{global}  transformations each $\alpha^a$ is
a constant.  In \underbar{local} transformations, the parameters $\alpha^a$
depend on space-time, $\alpha^a = \alpha^a (t, \vec{x})$.
\b
The explicit form of the generators depends on the representation of the field
to which $D_\mu$ is applied. Examples are: $T^a=\tau^a/2$
for fields that are doublets of $SU(2)$, where $\tau^a$,
$a=1,\th 2,\th 3$, are the Pauli matrices;
 $T^a=\lambda^a/2$ for fields in the representation [3] of $SU(3)$,
 where $\lambda^a$, $a=1,\th \cdots ,\th 8$, are the
Gell-Mann matrices. $SU(n)$, the
group of unitary $\rm n\times n$ matrices with determinant equal to one, has
$\left( n^2\th -\th 1\right)$ generators $T^a$. They obey the commutation
relations
$$\left[ T^a,\th T^b\right] = i\th C_{abc}\th T^c~, \eqno {(24)}$$
(sum on c) where
$C_{abc}$ are numbers called structure constants. For example, in $SU(2)$
$$\left[T^a,\th T^b\right] = i\th \epsilon_{abc}\th T^c~, \eqno{(25)}$$
\no where $\epsilon_{abc}$ is the completely antisymmetric tensor with three
indices.
\b
The commutation relations are characteristic of the group, and not of
the
particular representation. For example the generators in the representation [2]
of SU(2)  are
$$T^a = {{\tau^a}\over{2}} = \left({
{{1}\over{2}}\left[\matrix{0&1\cr 1&0\cr}\right] ,\ {{1}\over{2}}\left[
\matrix{0&-i\cr i&0\cr}\right] ,\ {{1}\over{2}}\left[\matrix{1&0\cr 0& -1\cr}
\right]}\right)~,\eqno{(26)}$$
\no and in the representation [3] of SU(3) they are
$$T^a = \left({{{1}\over{\sqrt{2}}}\left[\matrix{0&1&0\cr 1&0&1\cr
0&1&0\cr}\right] ,\ {{1}\over{\sqrt{2}}}\left[\matrix{0&-i&0\cr i&0&-i\cr
0&i&0\cr}\right] ,\ \left[\matrix{1&0&0\cr 0&0&0\cr 0&0&-1\cr}\right]}\right).
\eqno{(27)}$$
\no In both cases the generators fulfill the relations in Eq. (25). In order to
obey the same commutation relations, the generators in different
representations have different normalizations, i.e.
$$Tr\th \left(T^a\th T^b\right)\th =\th r\delta^{ab}~,\eqno{(28)}$$
\no where $r$ is a representation dependent number (this is important for
GUT's). For example, for $SU(2)$, $r_{[2]} = 1/2$, $r_{[3]} = 2$. Usually $r$
is chosen to be $r=1/2$ in the fundamental representation [n] of $SU(n)$.
\b
The number of simultaneously diagonalizable generators, i.e. generators that
commute with each other, is called the rank of a group. The set of these
generators corresponds to the set of quantum numbers that can be defined for
each particle. The rank of $SU(n)$ is (n$-$1). Thus, $SU(2)$ has only one
diagonal
generator, $T_3$, and $SU(3)$ has two, usually chosen to be $T_3$ and $T_8$.
The rank of the standard model is, thus, $2+1+1 =4$.
\b
The gauge invariant Lagrangian derived from Eq. (21) is
$${\cal L}_{\rm gauge} = {\cal L}\left(\chi_\alpha ,\th
D_\mu\chi_\alpha\right)\th +\th {\cal L}_{\rm g.b.}~.\eqno{(29)}$$
\no The last term is the Lagrangian of the gauge bosons
$$\eqalignno{
{\cal L}_{\rm g.b.} =& -\th {{1}\over{4}}\th \sum_{a}\th F^a_{\mu\nu}\th
F^{\mu\nu a}~, & {(30)}\cr
\noalign{\hbox{\rm with}}\cr
F^a_{\mu\nu} =~& \partial_\mu\th A^a_\nu\th -\th \partial_\nu\th
A^a_\mu\th +  {\rm g}~C_{abc}\th A^b_\mu\th A^c_\nu~. & (31) \cr}$$
\no For a U(1) group, $C_{abc}=0$. Thus, only for a $U(1)$ group there are no
three-gauge-boson and four-gauge-boson couplings. Gluons and the weak gauge
bosons do couple with themselves, while photons do not.

In addition to the usual term $\sum_{a}\th F^a_{\mu\nu }\th F^{\mu\nu a }$ in
Eq. (30), also the term \break $\sum_a\th F^a_{\mu\nu}\th
 \widetilde{F}^{\mu\nu a}$,
 with $\widetilde{F}^{\mu\nu a} =
(1/2)\epsilon^{\mu\nu\alpha\beta}\th F^a_{\alpha\beta}$, is
invariant under gauge transformations.
 Thus, except for
U$_Y$(1) (for which $\widetilde{F}^{\mu\nu} = 0$) we should add to the
Lagrangian in Eq. (29) a term
$${{{\rm g}~{\theta}}\over{32\pi^2}}\th \sum_{a}\th F^a_{\mu\nu }\th
\widetilde{F}^{\mu\nu a} \eqno{(32)}$$
\no for each group of the S.M.
This term can be written as a total
divergence, thus its contribution to the
action $S = \int  {\cal{L}}~ d^4x$, can be transformed into a surface
integral at
$\vert x\vert \to \infty$. This can be chosen to be zero (with gauge fields
going to zero sufficiently fast as $\vert x\vert\to\infty$) unless
non perturbative field configurations are taken into account. The existence of
non-perturbative tunnelling amplitudes between different vacua
(with different
topological properties) yields a non-zero contribution of the term in Eq. (32)
to the action. The tunnelling amplitudes are of
order $exp\left( -4\pi /{\rm
g}^2\right)$. This factor is of order
$10^{-172}$ for SU(2).
Thus, the term in Eq. (32) is irrelevant for $SU(2)$. However, it cannot be
avoided for $SU(3)$, where $g_3$ can be of $O(1)$.
This term, present in the QCD Lagrangian, violates P and CP, which implies an
experimental upper bound $\theta_{\rm QCD}<10^{-9}$ for the arbitrary parameter
$\theta_{\rm QCD}$. Why is $\theta$ so small is an open problem in QCD, ``the
strong CP-problem'', that may provide a window towards physics beyond the S.M.
For example, it may require the existence of a global spontaneously broken
quasi symmetry, the U(1) of Peccei and Queen$^{5}$ .  We will not deal with
this problem in these lectures.
\b
The globally invariant Lagrangian for the fermions of the S.M.,
$$ {\cal{L}} = \sum_f\th (\bar{f}_L\gamma^\mu \partial_\mu\th f_L+
\bar{f}_R\th
\gamma^\mu\partial_\mu\th f_R)~,\eqno{(33)}$$
becomes a gauge invariant Lagrangian for the fermions, after replacing
$\partial_\mu\to D_\mu$,
$$\eqalign{
{\cal{L}}_f =&\sum_f\th \bar{f}_L\th \gamma^\mu\left(\partial_\mu\th
-\th i{\rm
g}_2\sum_i\th \left({{\tau^i}\over{2}}\right)\th W^i_\mu\th -\th i{\rm g_1}
Y_L\th B_\mu\right) f_L\cr
+& \sum_f\th f_R\th \gamma^\mu \left(\partial_\mu\th -\th i{\rm g_1}Y_R\th
B_\mu\right)\th f_R\cr
+& \sum_q\th \bar{q}_L\th \gamma^\mu\left(-i{\rm g_3}\sum_a\th
{{\lambda^a}\over{2}}G^a_{\mu}\right)\th q_L\cr
+& \sum_q\th \bar{q}_R\th \gamma^\mu\left(-i{\rm g_3}\th \sum_a\th
{{\lambda^a}\over{2}}G^a_{\mu}\right)\th q_R~.\cr} \eqno{(34)}$$
\no Here $g_1,\th g_2,\th g_3$ are the coupling constants of U$_Y$(1),
 SU$_L$(2)
and SU$_C$(3) respectively, $i=1,\th 2,\th 3$, and $a=1,\th \cdots\th ,\th 8$,
label the generators of SU$_L$(2)
and SU$_C$(3) respectively, $f$
indicates all fermions, while $q$ indicates only the quarks. The last
 two terms
correspond to $SU_C(3)$. Notice that the $SU_C(3)$ couplings are the
 same for the left- and right-handed quarks.
\b
\no{\it 2.3. ~Global Symmetries}
\s
In the absence of mass terms $m_f \bar{f}_L\th f_R$ in the Lagrangian,
$f_L$ and $f_R$
are not mixed. Thus, the $N_L$ left-handed fermions
with equal quantum numbers can
be rotated among themselves by unitarity global $U_L\left( N_L\right)$
transformations $U$, since
${\bar{f}}_L\th
U^{-1}\th Uf_L = {\bar{f}}_L f_L$. Also
the $ N_R$  massless fields
$f_R$ with equal quantum numbers can
be rotated among themselves by unitarity global $U_R \left( N_R\right)$
transformations.
Thus, in general, the global symmetry of a gauge fermion Lagrangian with
zero masses is
$$U_L\left( N_L\right)\times U_R\left( N_R\right) = SU_L\left( N_L\right)
\times
U_L(1)\times SU_R(N_R)\times U_R(1)~.\eqno {(35)}$$
\no The generators of the groups acting only on $f_L$ contain the projector
$P_L$, those acting only on $f_R$ contain $P_R$,
$$T^a_L = T^a\th {{\left( 1\th -\th \gamma_5\right)}\over{2}},\qquad T^a_R
 = T^a \th
{{\left( 1\th +\th \gamma_5\right)}\over{2}}~. \eqno{(36)}$$
\no The vector and axial-vector diagonal groups have, respectively, the
generators
$$T^a_V = T^a_L\th +\th T^a_R\th =\th T^a,\qquad T^a_A\th =\th T^a_R\th -\th
T^a_L\th =\th T^a\gamma_5 ~.\eqno{(37)}$$
\no In order to obtain these diagonal generators we need $N_L=N_R=N$.
In the actual world,  the approximation $m_u =
m_d = 0$ yields predictions which are good to a few percent, thus
the QCD Lagrangian (neglecting the electroweak interactions) has
an approximate global  symmetry with $N=2$, $U_L(2)\times U_R(2)$.
 If also $m_s=0$ is assumed, the predictions are only good to a few
10\th\%, since the $s$ quark is heavier than the $u$ or $d$, and $N=3$.

The
four global symmetries of massless fermions are very different. $U_{V}(1)$
is a
conserved symmetry, the baryon number in QCD. $U_{A}(1)$ is instead,
anomalous,
i.e. not a true symmetry once quantum corrections are taken into account. The
$SU_{V}(N)$ symmetry shows explicitly in degenerate multiplets of particles.
This is the strong-isospin $SU(2)$ symmetry  (that shows in doublet or
triplet
multiplets of almost degenerate particles, such as $(p,\th n)$, $(K^+K^0)$,
$(\pi^+\th\pi^-\th\pi^0$), etc.) or the flavour $SU(3)$  symmetry  of the
``eightfold way"$^{6}$ (that shows in octets and ten-plets of almost
degenerate particles). The axial $SU_A(N)$  is, instead, spontaneously broken.
It is believed that the light quarks condense in the vacuum,
$$\langle
0\vert\bar{u}_L\th \bar{u}_R\vert 0 \rangle\th =\th \langle 0\vert
 \bar{d}_L\th
d_R\vert 0\rangle\th =\th \cdots \th \neq 0~~,\eqno{(38)}$$
 at an energy scale close to $\Lambda_{QCD}$, where the strong coupling
 constant $\alpha_s$ becomes of $O(1)$ (see Fig. 1.)
\vglue 2.5truein
\no Fig. 1.  Behavior of the strong coupling constant $\alpha_s
(q^2) ~vs.~ q^2$, the square momentum transfer, showing the regions of
 asymptotic freedom, chiral symmetry
breaking and confinement.
\bb
\no The quark bilinears in Eq. (38) are not $SU_A(N)$ singlets, thus,
 their non-zero
V.E.V. break chiral symmetry spontaneously in to the vector diagonal group,
 $SU_L(N)\times SU_R(N)\to SU_V(N)$. The spontaneous breaking of a global
symmetry is signaled by the appearance of a zero mass boson, called a Nambu-
Goldstone boson $^7$,  for each broken generator. For the broken symmetry
$SU_A$(2) of the QCD Lagrangian, these bosons
are the three pions, $\pi^+,\th\pi^0,\th\pi^-$. Because the light quark
 masses
are actually non-zero, $SU_A(N)$ is only an approximate symmetry, whose
spontaneous violation yields ``quasi" Goldstone bosons, i.e. bosons with a
small mass (see below). Thus, the pion mass $m_\pi$ is non-zero, but it is
still much smaller
than all other hadronic masses.
\b
\no {\it 2.4. ~Spontaneous Symmetry Breaking}
\s
A spontaneously broken global symmetry is a symmetry of the Lagrangian that is
not respected by the vacuum state. Let us see explicitly the appearance of zero
mass bosons in a simple example. Consider a complex scalar field $\phi =
(\phi_R\th +\th i\phi_I)/ \sqrt{2}$ and a Lagrangian that is invariant under
 $U(1)$
transformations of $\phi$, i.e. under the multiplication of $\phi$ by a
phase
$$\phi\to\phi^\prime\th =\th e^{i\alpha}\phi~.\eqno{(39)}$$
\no The most general renormalizable Lagrangian, invariant under the
transformation in Eq. (39) is
$$\eqalignno{
{\cal L} = &~ \partial^\mu\th\phi^{\dag}\th \partial_\mu\phi\th -\th
V\left(\phi^{\dag}\th
\phi\right)~, & (40)\cr
\noalign{\hbox{\rm with}}
V\left(\phi^{\dag}\th \phi\right) =& -m^2\th \phi^2\th\phi\th
+\th\lambda\left(\phi^{\dag}\th\phi\right)^2~. & (41)\cr}$$
\no While the condition of having $V$ bounded from below for large values
 of
$\phi$ imposes $\lambda > 0$, the quadratic term could have a positive or
negative coefficient. If $m^2<0$, there would be only one minimum of $V$, at
$\phi = 0$, and $U(1)$ would be a manifest symmetry. We are interested in the
case $m^2 > 0$. The minimum of the potential corresponds to
$${\phi^{\dag}\th
\phi} = {{v^2} \over{2}} = {{m^2}\over{2\lambda}}~.\eqno{(42)}$$
\no Any point in this ``orbit'' of degenerate minima in
the plane $\phi ,\th \phi^{\dag}$ is an equally good minimum (See Fig. 2). The
system will choose one of these points and we choose the axis of real values of
$\phi$ passing through it. Thus, without loss of generality the vacuum is at
$$\langle \phi\rangle = {{\langle \phi_R \rangle} \over{\sqrt{2}}}
= {{v} \over{\sqrt{2}}}~. \eqno{(43)}$$
\no Notice that the curvature along the imaginary axis is zero, since
this is the curvature of the  orbit of degenerate minima,
i.e.
$$m_{\phi_I} = 0~. \eqno{(44)}$$
$\phi_{I}$ is, thus, the Nambu- Goldstone$^7$ boson in this example.
\vglue 2.8truein
\no Fig. 2. The potential in Eq. (41), for $m^2 > 0$
\bb
\no Alternatively, we could parameterize $\phi$ in ``polar coordinates''
(this is
called the ``unitary gauge''),
$$\phi={{1}\over{\sqrt{2}}}\left[ v + \rho(x) \right] e^{i\xi (x)/v}~.
\eqno{(45)}$$
 By replacing Eq. (45) in the potential, Eq. (41), we can explicitly find
 that
$$m_\xi = 0~ .\eqno{(46)}$$
\no Thus, $\xi (x)$ is the Nambu- Goldstone boson in this case.
 Notice that
for small values of $\xi (x)$ we have $\phi\simeq {{1}\over{\sqrt{2}}}\left( v
+ \rho (x) + i\xi (x)\right)$ and $\xi (x) = \phi_I (x)$.

If the global symmetry is only an approximate  symmetry, i.e. if there are
small terms in the Lagrangian that do not respect the global symmetry, the
orbit of minima is no longer flat but  slightly deformed. Thus the curvature of
the potential in a direction tangent to the orbit of minima is non-zero. This
curvature is the mass of the Nambu- Goldstone boson. Instead of a zero mass
Nambu- Goldstone boson, we have a ``pseudo'' or ``quasi'' Nambu- Goldstone
boson with a small mass.  Pions are quasi-Goldstone bosons, as we mentioned
above.
\b
When a global spontaneously broken symmetry is gauged, something amazing
happens. The Nambu- Goldstone bosons disappear as  physical particles, they are
incorporated into (``eaten up'' by) the gauge bosons that become massive. This
is the ``Higgs effect''$^{8}$ . Let us see it happening in our $U(1)$ example.
Let us replace
$$\partial_\mu\to D_\mu = \partial_\mu\th -\th i{\rm g}\th A_\mu\eqno{(47)}$$
\no in the globally invariant Lagrangian in Eq. (40). Thus,
$$D_\mu\th \phi= {{e^{i\xi /v}}\over{\sqrt{2}}}\th \left[\partial_\mu\rho\th
-\th i\left( v+\rho\right)\th {\rm g}\left( A_\mu\th -\th
{{\partial_\mu\xi}\over{v{\rm g}}}\right)\right]~.\eqno{(48)}$$
\no Define the new gauge field
$$B_\mu = A_\mu\th -\th {{\partial_\mu\xi}\over{v{\rm g}}}~. \eqno{(49)}$$
\no Thus, the term $(D_\mu\phi)^{\dag}\th D_\mu\phi$ contains a term
$$-i^2\th {\rm g}^2\th v^2\th B_\mu\th B^\mu~. \eqno{(50)}$$
\no and $\xi (x)$ is entirely eliminated from the Lagrangian. From Eq. (50) we
see that $B_\mu$ is massive
$$m_B = {\rm g}v~.\eqno{(51)}$$
\no In Eq. (49) we see that the massive gauge boson $B_\mu$ incorporates
$\partial_\mu\xi$ (as a longitudinal component, besides the two transverse
components in $A_\mu$).
\b
Let us return to the three pions, $\pi^+,\th\pi^0,\th \pi^-$, as the
pseudo- Nambu- Goldstone bosons of the global axial $SU_L(2)$ approximate
symmetry of the QCD Lagrangian. This symmetry is spontaneously broken by the
non-zero V.E.V. of quark bilinear operators in Eq. (38), of the form
$$\langle 0\vert \bar{q}_L\th q_R\vert 0\rangle~, \eqno{(52)}$$
\no since (${\bar{q}}_L\th q_R$) is not
invariant under  axial transformations
$$q\to e^{-i \alpha_a  \gamma_5 T^a} q~,  \eqno{(53)}$$
\no as we will now prove. An axial transformation on the Dirac field $q$,
corresponds  to transforming $q_L$ and $q_R$ with
opposite phases,
$$q_R\to e^{i\alpha_aT^a}\th q_R~, \quad q_L\to e^{-i\alpha_aT^a}\th
q_L~.\eqno{(54)}$$
\no It is easy to see that it is so, by considering only infinitesimal
transformations, since any finite transformation can be
obtained by a succession of infinitesimal ones. Let us, thus, consider
 axial transformations by infinitesimal
phases $\left( \delta\alpha\right)_a$ and see that right and left-handed
fields are rotated by opposite phases:
$$\eqalign{
q\to& q+ i(\delta\alpha)_a\th \gamma_5\th T^a\th q = \left( q_R+q_L\right)
+ i(\delta\alpha )_a\th \gamma_5\th
\left[{{1+\gamma_5}\over{2}}+{{1\th -\th\gamma_5}\over{2}}\right] q\cr
& = \left( q_R + i (\delta\alpha )_a\th T^a\th q_R\right)\th +\th
\left( q_L\th
-\th i(\delta\alpha)_a\th T^a\th q_L\right)~~.\cr}\eqno{(55)}$$
\no  Now, using Eq. (54) we see
that $\bar q_L\th q_R$ is not invariant under axial rotations,
$$\bar{q}_L q_R\to \overline{\left( e^{-i\alpha_a\th T^a}q_L\right)}
\th \left(
e^{i\alpha_a\th T^a}q_R\right) = \bar{q}_L e^{2i\alpha_ a\th
T^a}q_R~.\eqno{(56)}$$

  We also know that in the S.M., $(\bar{q}_Lq_R$) is \underbar{not} invariant
under electroweak transformations. Thus, the non-zero V.E.V.'s in Eq. (38) not
only break the global chiral  symmetry spontaneously,  but they also break
the gauge symmetry $SU_L(2)\times U_Y(1)$ spontaneously. If there would be
 no other
electroweak breaking mechanism (i.e. no Higgs field), the $\bar{q}_L q_R$
condensates would give mass to the $W^\pm$ and $Z$, at their energy scale
$$m_{W^{\pm},\th Z}\simeq \wedge_{\rm QCD} \simeq 100\th {\rm
MeV}~.\eqno{(57)}$$
\no and the three pions $\pi^+,\th \pi^0,\th \pi^-$ would be ``eaten up'' by
the weak gauge bosons of the same charge. This mechanism for
electroweak  dynamical symmetry
breaking is at the basis of Technicolour models.
The breaking is called ``dynamical'' because the condensates form as the result
of interactions as their coupling constant becomes of order one, in this case,
colour interactions (see Fig. 1.).
\b
\no {\it 2.5. ~Spontaneous Electroweak Breaking in the Standard Model}
\s
In the S.M., the electroweak spontaneous breaking is done by introducing a
fundamental complex colour singlet scalar $\phi$, that is a doublet of
SU$_L$(2) and carries weak hypercharge $Y_\phi = -1/2$. Both the field $\phi$
and its conjugate, ${\tilde{\phi}} = i \tau^2 \phi^{\ast}$, are used
$$\phi= \left(\matrix{\phi^0\cr\phi^-\cr}\right)~,~~\qquad{\tilde{\phi}} =
\left(\matrix{\phi^+\cr{-\bar{\phi}}^0\cr}\right)~.\eqno{(58)}$$
\no These fields are coupled in a gauge invariant way,
$${\cal L}_{\rm Higgs} = (D^\mu\th \phi)^{\dag}\th D_\mu\th \phi\th -\th
V (\phi^{\dag}\th \phi)~,\eqno{(59)}$$
\no with an ad-hoc potential as in Eq. (41) (where now $\phi$ is a doublet),
that has a non-zero V.E.V. given by Eq. (42). Since $\phi^{\dag}\th \phi=
{\tilde{\phi}}^{\dag}\th {\tilde{\phi}}$, it is equivalent to write the
Lagrangian in terms of $\phi$ or ${\tilde{\phi}}$.  The component of
$\phi$ with
non-zero V.E.V. is a singlet under a $U(1)$ symmetry. This
$U(1)$  is, therefore, preserved as an unbroken symmetry, and it is
identified with electromagnetism, $U_{\rm em}(1)$. Thus,
$$\langle \phi\rangle = \left(\matrix{\langle \phi^0\rangle\cr
\langle\phi^-\rangle\cr}\right) =\ \left(\matrix{v/\sqrt{2}\cr  0 \cr}\right)
\eqno{(60)}$$
where $v = (m^2/\lambda)^{1/2}$, (see Eq. 42) given the parameters of the
potential in Eq. (41).  Since $\phi^0$ has zero electric charge, $< \phi^0 >
\not=0$ does not break $U_{\rm em}~(1)$.  The photon remains massless, and the
Higgs mechanism gives masses to the other gauge bosons,
$$ m_{W^{\pm}}^2 = {g_2^2 v^2\over 4}~,~~~ m_{Z^0} = (g^2_1 + g^2_2)
{v^2\over 4} = {m_W^2\over \cos^2 \theta_W}~. \eqno{(61)}$$
The spontaneous symmetry breaking also gives origin to all fermion masses
through the Yukawa couplings,
$${\cal{L}}_{\rm Yukawa} = \sum_{\alpha, \beta} \th \left[
(\lambda_{u})_{\alpha \beta}~\th {\bar{q}}_{L \alpha}\th \phi \th u_{R \beta}
+ (\lambda_d)_{\alpha \beta}~ {\bar{q}}_{L \alpha} {\tilde{\phi}} d_{R\beta
+} + (\lambda_{\ell})_{\alpha \beta} ~{\bar{\ell}}_{L \alpha} {\tilde{\phi}}
e_{R \beta} \right]~, \eqno{(62)}$$
where $\alpha, \beta = 1,2,3$ are generation indices and $(\lambda_f)_{\alpha
\beta}$ are coupling constants.  The fermion mass matrices are
$$ (m_f)_{\alpha \beta} = (\lambda_f)_{\alpha \beta} {v\over
{\sqrt{2}}}~. \eqno(63)$$
In Eqs. (61) and (63) we see that
$v$ fixes the scale of all the masses in the S.M.  From Eq. (61),
given the masses of the $W$ and $Z$, fields, we obtain
$$ v~ \simeq~ 250 GeV~. \eqno(64)$$
\b
\no{\bf 3. ~Unsatisfactory Features of the Standard Model}
\b
Let us enumerate those features most used against taking the S.M. as the
ultimate elementary particle theory.
\b
\no {\it 3.1. ~Too Many Free Parameters}
\s
There are 19 parameters, if
$\nu's$ are massless.  They are:
\b
\item{-} 3 coupling constants $~~g_1, g_2, g_3$ or $~~\alpha$,
$\theta_{W}$, $\Lambda_{QCD};$
\b
\item{-} 2 parameters of the Higgs potential $v$ and $\lambda$, or
$m_{Z^0}$ and $m_{\rm Higgs}$;
\b
\item{-}9 fermion masses: $m_e, m_{\mu}, m_{\tau}$; $m_u, m_c, m_t$;
$m_d, m_s, m_b$;
\b
\item{-} 3 mixing angles and 1 phase in the Kobayashi-Maskawa$^{9}$
matrix;
\b
\item{-} the vacuum parameter of $QCD$, $\theta_{QCD}$.
\b
There are several more if $m_\nu \not=0$.  But, how many parameters are
acceptable in a fundamental theory?  This is more a philosophy question
than a physics question.  Still, the goal is to explain as much as
possible with the least number of parameters.
\b
\no {\it 3.2. ~The Mass Scale $v$ is Not ``Natural"}:
\s
A naturally small quantity is a quantity whose small value has a
symmetry reason, i.e.,  the symmetry of the system would increase if it were
zero.  We said that the only mass scale in the S.M. is $v$.  This is
true in the sense that all masses are proportional to $v$.  However,
there is another relevant scale, a scale $\Lambda >> v$.  The S.M. can
\underbar{not} be a good theory for arbitrarily large energies, thus
there must be a cut-off energy scale, $\Lambda$.  Since gravity is not
included in the S.M., there is for sure a cut-off at $\Lambda \simeq
M_{\rm Planck} \simeq 10^{19}$ GeV.  If there is new physics below the
quantum gravity scale (such as GUT's, or Technicolour, etc.),
$\Lambda$ may be lower.  Thus,
$$ v < < \Lambda \eqno(65)$$
leads to two questions:
\s
\item{i)} Why does $v$ have the value it has?
\s
\item{ii)} Assuming we give $v$ the value it has at tree level, is Eq.
(65) a stable condition?
\no In fact, one expects $v^2 \to v^2 + \Lambda^2$, due to radiative
corrections.  This is because radiative corrections to the Higgs mass, $m
= v/\sqrt{\lambda}$, from gauge (and gravitational) interactions are
quadratically divergent
$$ \delta m^2~\simeq~g^2 \int^{\Lambda}_0 {d^4 k\over (2 \pi)^4
k^2}~\simeq~ O(\alpha^2) \Lambda^2~, \eqno(66)$$
where $\alpha=  g^2 /4 \pi$.Thus $\delta v^2~\sim \Lambda^2 \lambda$. Call
$v_0$ the tree level value, thus
$$ v^2 = v^2_0 - \lambda \Lambda^2~. \eqno(67)$$
\no In order to obtain $v \simeq 250~$GeV we would use a trick similar to
that used in renormalization, namely, we assign to the bare mass $m^2_0 =
v_0^2/\lambda$ the value necessary to obtain the result we want,
$$ \lambda (m^2_0 - \delta m^2) = \lambda \left[ O (\Lambda^2) - O
(\Lambda^2) \right] \simeq (250~GeV^2)~.\eqno(68)$$
Assuming $\Lambda \simeq  10^{16}$ GeV, a typical GUT scale, we assume
in Eq. (68) that two quantities of order $10^{32}$~almost cancel
each other to produce a result of order $10^4$.  This is a fine-tuning by
28 orders of magnitude!
\b
\no Two solutions to this problem have been proposed.  One of them
consists of
eliminating the Higgs field altogether, relying on a dynamical mechanism to
break spontaneously the electroweak symmetry.  The only mechanism of this type
we know is the chiral symmetry breaking by quark bilinear condensates $<
\bar{q} q >$ due to colour interaction.  We have seen that this mechanism
 would
produce a mass of $O (\Lambda_{QCD})$ for the $W^{\pm}$ and the $Z$.
Thus, one could consider a repetition of this mechanism at a larger
scale $\Lambda_{TC}$, where condensates $< \bar{F} F>$ of new fermions,
the ``Techni-fermions", would form due to new interactions, ``Technicolour"
interactions$^{10}$.

The second proposal cures the problem ii) by introducing a fermion-boson
symmetry, a ``supersymmetry" $^{11}$.  Then, fermion loops contribute to
radiative corrections exactly the same as bosons do, but with the
opposite sign (due to Fermi-statistics).  Thus
$$ \delta m^2_{\rm Fermions}~\simeq~ - g^2_y \int^{\Lambda} {d^2 k\over
(2 \pi)^4 k^2}~\simeq~ - O (\alpha_y) \Lambda^2~. \eqno(69)$$
where $g_y$ are Yukawa-type couplings, that, due to supersymmetry
coincide with gauge couplings $g$.  The cancellation $\delta m^2_{\rm
bosons} - \delta m^2_{\rm fermions} = 0$, would be exact with an exact
supersymmetry that guarantees equal masses for bosons and femions, $m_b =
m_f$.  However, supersymmetry must be broken, since we have not observed
the supersymmetric partners of the standard particles.  Thus
$$ \delta m^2 = O (\alpha) \vert m^2_b - m_f^2 \vert. \eqno(70)$$
If, the mass differences are
$$ \vert m_b - m_f \vert \leq O( 1~ TeV)~~, \eqno(71)$$
then $v$ is stable.  Eq. (70) means that we should find supersymmetric
partners with $m \leq O (1~ TeV)$, if supersymmetry answers the question
ii) above.  Notice that supersymmetry does not necessarily explain why
$v$ has the value it has in the first place.
\b
\no {\it 3.3. ~The Interactions are Not Unified}
\s
The gauge group $SU_C (3)\times SU_L(2) \times U_Y(1)$ has three factors, thus
three independent couplings constants.  Can the idea of unification contained
in the electroweak unification go further?  Grand Unification$^{12,~4}$
proposes the existence of a simple symmetry group G that breaks
spontaneously into the S.M.
group at a large energy scale $\Lambda_{GUT}$.  Supporting evidence for GU
comes from renormalizations: the three running coupling constants tend to
become less different as the energy scale increases.  Since $\Lambda_{GUT} > >
v$, GUT's suffer from the problem of naturality of $v$, or ``gauge hierarchy
problem".  Supersymmetry may be invoked to solve this problem, yielding
 supersymmetric GUT's $^{13}$.
\b
\vfill \eject
\no {\it 3.4. ~There is No Reason for the Existence of Three Families}
\s
Proposals to explain the existence of three families and their mass
hierarchy have been made following one of the two guiding principles
mentioned above, i.e., either more symmetry or another layer of
compositeness.  ``Horizontal" or ``inter-familiar" symmetries have the
three fermions of equal quantum numbers as members of a multiplet.  The
horizontal symmetry   group $G_H$ is independent   of the ``vertical"  or
``intra-familiar" group $G_V$, so that the total  gauge group is $G_V \times
G_H$. The names ``vertical" and ``horizontal" stem from our habit to write the
families in columns. $G_V$ could be the S.M.   or a vertical G.U. group.  $G_H
$ could be $SU(2)$ or $SU(3)$ $^{14}$, for example. In these models the
inter-familiar mass hierarchy is due to the spontaneous symmetry breaking of
$G_H$.  Realistic mass matrices can only be obtained with very complicated
schemes.  Also ``vertical-horizontal" Grand Unifications have been proposed
with large unification groups, such as $SO(16)$ or $SO(18)$ in rather baroque
models $^{15}$.  We will not talk about this approach here.

In composite models, quarks and leptons are bound states of
sub-components, that may have different names, one of them being ``preons".
In these models the family replication at higher mass scales is
explained by the existence of excited levels of the bound states of
preons.
\b
\no {\it 3.5. ~Gravity is Not Included}
\s
General relativity can be formulated as a classical field theory,
but attempts
to quantize it yield a non-renormalizable theory.  The hope is to
unify gravity
with other forces in such a way that the infinities arising in different
sectors cancel among themselves, yielding a combined finite or
renormalizable
theory.  The hope is that also in the same theory there will be an explanation
for why the cosmological constant is so small.  The best candidate at present
for such a unification is provided by local-supersymmetry, that includes
gravity automatically, and is thus, called ``supergravity" $^{16}$.
Supergravity is one of the features of superstring models.
\b
\no {\it 3.6. ~The S.M. Does Not Provide Dark Matter Particle Candidates}
\s
Dark matter (D.M.), the most abundant form of matter in the universe,
 constitutes 90 to 99\% of the total mass.  Its density in units of the
critical density, $\Omega_{\rm D.M.}$,  is known to be between $0.2$ and $1$,
while the luminous regions in the universe account only for a fraction
$\Omega_{\rm lum}\simeq 0.01$ of the critical density $^{17}$.  The nature of
the D.M. is one of the most important open questions in physics and
astrophysics.  It may consist of nucleonic matter in the form of very low mass
stars, the brown dwarfs, too small to shine, or remnants of dead stars, or
primordial black holes.  There are current searches trying to detect some of
these objects in the dark halo of our galaxy $^{18}$.  The upper bound on
nucleonic matter coming from nucleosynthesis in the early universe, makes it
very difficult to have much more than $\Omega \simeq 0.1$ in nucleonic
D.M.$^{19}$.  This bound strongly suggests that, even if non-shining objects
made of nucleonic matter may account for the dark halos of galaxies, nucleonic
D.M. may not be enough to account for the larger amount of D.M. bound to larger
scales (clusters and super-clusters of galaxies).  Thus, if one believes in
nucleosynthesis bounds, the bulk of the D.M. should consist of more exotic
types of matter, such as new neutral massive elementary particles.  The only
possible such particles contained in the S.M. are neutrinos, but, in the S.M.
they are massless.  Thus, to provide for D.M. particle candidates the S.M. must
be extended.  The good news is that most extensions of the S.M. easily provide
potential D.M. candidates.  Let us see why.
\b
To give the right mass to the light neutrinos seems the
easiest way to get a D.M. candidate.  The present relic abundance in
number of the three neutrino species of the S.M. is fixed by their
thermal history.  Thus, if the sum of the masses of all three neutrinos
is $\simeq  10 -30$ eV they will account for $\Omega \simeq 1~^{17}$.
\b

Most extensions of the S.M. predict neutral heavy stable particles with
masses in the GeV-TeV range and interactions of the weak order.  The
amazing fact is that particles with these characteristics have
$\Omega \simeq 1$ and are, thus, ``natural" D.M. candidates.  These
particles are generally called WIMP's, Weakly Interacting Massive
Particles (by opposition, astrophysical D.M. candidates are called
MACHO's, Massive Compact Halo Objects).  For WIMP's the relic abundance
turns out to be $^{17}$
$$ \Omega h^2 \simeq 1 \cdot 10^{-37} cm^2 < \sigma_a v >^{-1}~,
\eqno(72)$$
where $h = 0.4 - 1$ is the Hubble constant in units of 100 Km/Mpc sec.,
 and $<\sigma_a v >$ is the thermal average of the annihilation cross
 section by the
relative velocity of the WIMP's.  Eq. (72) holds in most models, even if
it can
be modified by  a non-zero cosmic asymmetry in WIMP - anti WIMP number,
 or by
generating entropy (mainly photons) after WIMP's annihilate.

Using order of magnitude and dimensional arguments on the annihilation cross
section, it is easy to
to determine the range of masses for which WIMP's could have the abundance
adequate to be D.M. candidates.  For non-relativistic particles lighter than
the $Z$ and $W$ weak gauge bosons (or other exchanged particles of similar
mass), dimensional arguments yield  roughly $\sigma_a \simeq G^2_F~ m^2~
N_a$, for the annihilation cross section.  Here $G_F$ is the Fermi coupling
constant, $m$ is the mass of the WIMP, and $N_a$ is the number of annihilation
channels. Thus, $\sigma_a \simeq 0.4~N_a~ 10^{-37} cm^2~(m/1GeV)^2$ and
$$ \Omega h^2 \simeq  O(1-10)(m/1~GeV)^{-2}~, \eqno(73)$$
the density decreases with increasing mass.
 WIMP's with these interactions must have masses of the order of $1$ to
$100$ GeV to account for $\Omega_{\rm D.M.}$.  WIMP's much heavier than the
weak gauge bosons (or any other exchanged particles) have ``electromagnetic
like" cross sections (since the mass of the exchanged particle can be
neglected).  Dimensional arguments yield, in this case,
$ \sigma_a\simeq \alpha^2 N_a/m^2\simeq 0.2~N_a~ 10^{-37} cm^2~
(m/1~TeV)^{-2}$, where $\alpha$ is the electromagnetic coupling constant.
Therefore, the density increases with increasing mass,
$$ \Omega h^2 \simeq O(1 - 10) (m/1~TeV)^2~. \eqno(74)$$
  D.M. candidates with these
interactions must have masses of the order of $1$ to $100~TeV$.

Results from LEP, combined with other bounds on D.M. have rejected
several of the WIMP D.M. candidates proposed up to now$^{20}$, but many
remain viable.  One of the preferred D.M. candidates is still the LSP,
the lightest supersymmetric particle, in supersymmetric models, as we
will see later.
\b
\no {\it 3.7. ~Additional Windows into Physics Beyond the Standard Model}
\s
Several other open problems may require physics beyond the S.M.  We have
mentioned one of them already, the strong CP problem$^5$.  Others are the
solar neutrino problem$^{21}$ and the atmospheric neutrino
problem$^{22}$.  Possible solutions may require a see-saw mechanism$^{23}$ to
obtain very small neutrino masses $(m_{\nu}~\simeq~ 10^{-3}~eV$ or
 $m_{\nu} \simeq
10^{-1}~eV)$.  We will return to these two problems later.  Still
another open
question is the origin of the baryon number (B) asymmetry in the universe,
i.e., the fact that we live in a world of matter only (with no antimatter).
Producing a B-asymmetry requires B and CP violation in a situation of
non-thermal equilibrium$^{17}$. It seems impossible to generate
the needed B-asymmetry just within the S.M. Trivial extensions
with more than one Higgs
field may work.  There are many possible solutions beyond the S.M.
\b
\no {\bf 4. ~Technicolour}
\b
This is a generic name for models$^{10}$ in which the
electroweak breaking is
assumed to happen as the spontaneous symmetry breaking of the axial
flavour symmetry in QCD.

In QCD quarks, $q$, are subjected to a gauge binding force, colour,
 associated
to a gauge symmetry $SU_C(3)$.  In Technicolour (T.C.) models,
Technifermions,
$F$, are subjected to a gauge binding force, Technicolour. Different T.C.
symmetry groups have been proposed, for example $SU_{ \rm TC}(N_{TC})$ or
$SO_{\rm TC} (N_{TC})$ with $N_{TC} \geq 2$.  In QCD, when the colour
interaction becomes strong enough, at a scale $\Lambda_{QCD}\simeq 200$ MeV,
condensates of a pair of quarks appear, like Cooper-pairs in the BCS theory of
superconductivity,
$$ < \bar{q} q >~\simeq~ O(\Lambda^3_{QCD} )~. \eqno(75)$$
Notice that the mass dimension of $(\bar{q} q)$ is 3 (3/2 for each
fermion). This explains Eq. (75), given that the only mass scale in QCD is
$\Lambda_{QCD}$.

 The same  is assumed to happen in T.C. models, in which
Technifermions condensate but at a scale $\Lambda_{TC} >> \Lambda_{QCD}$,
$$ < \bar{F} F >~\simeq~ O(\Lambda^3_{TC} )~. \eqno(76)$$

  As we mentioned above, in the
absence of any other electroweak symmetry breaking the three
pions would be
``eaten up" by the $W^{\pm}$ and $Z$, through the Higgs mechanism.  In this
case $m_{W^{ \pm},Z} \sim  f_{\pi}$, for example $ m_{W^{\pm}} = g_2 f_{\pi}/2
\simeq 30~MeV $, obviously much too low. To get  the actual value for
$m_{W^{\pm},Z}$, we would need $f_{\pi}$ to be  $\simeq 250 GeV$ instead of its
actual value of $93 MeV$. In Technicolour models the $W^{ \pm}$ and $Z$ bosons
``eat up" Technipions, $T \pi$, and acquire a mass
$$ m_{W^{\pm}, Z} \sim f_{T \pi}~, \eqno(77)$$
 in which $f_{T \pi}$ replaces $f_{ \pi}$. Therefore,
$f_{T \pi}$ should be of the order of the Higgs field V.E.V. $v$ in
the S.M., $f_{T \pi} \simeq v \simeq 250~GeV$. Since $\Lambda_{QCD}\simeq 2
f_{\pi}\simeq 200 MeV$, a similar relation is assumed for the T.C.
counterparts
$$ \Lambda_{TC}~\simeq~ 2 f_{T \pi}~\simeq~ 500~GeV~. \eqno(78)$$
Let us describe the simplest T.C. model$^{10}$.  The gauge group is,
$$ SU_{TC} (N_{TC}) \times SU_C(3) \times SU_L(2) \times U_Y (1)~, \eqno(79)$$
where $N_{TC} \geq 2$, the number of Technicolours, is left as
 a free parameter. The group
$SU_{TC}(N_{TC})$ yields an asymptotically free gauge interaction that
confines at $\Lambda_{TC}$.  The minimal Technifermion content is just
one generation of Techniquarks $Q$:
$$ (N_{TC}, 1, 2, 0):~ \pmatrix{U_L \cr D_L \cr}_{1,2, \cdots, N}~~~
\matrix{(N_{TC}, 1, 1, 1/2) &:~ U_{R~1,2,\cdots , N} \cr
(N_{TC}, 1, 1, -1/2) &:~ D_{R~1,2 \cdots, N} \cr}
\eqno(80)$$
Notice that with $Q = T_3 + Y$ we get $Q_U = 1/2, Q_D = -1/2$.  Thus the
sum of all electric
charges is zero, and the model is anomaly free.  If $m_U = m_D
= 0$ there is a flavour Techni-isospin global symmetry $SU_L(2) \times
SU_R(2)$.  The condensates
$$ < {\bar{U}}_L U_R > = < {\bar{U}}_R U_L > = < {\bar{D}}_L D_R > = <
{\bar{D}}_R D_L > \simeq O(\Lambda^3_{TC} ) \eqno(81)$$
break spontaneously the entire axial global $SU_A(2)$ and also break the
electroweak gauge symmetry $SU(2) \times U(1)~{\rm into}$ $U_{em} (1)$. The
Nambu- Goldstone bosons $ T \pi^{\pm} $ and $T\pi^0$ are eaten up by the
$W^{\pm}$ and the $Z^0$ respectively, so that
$$ m_{W^{\pm}} = {g_2\over 2}~ f_{T \pi^{\pm}}~,~~ m_{Z^0} = {g_2~ f_{T
\pi^0}\over 2 \cos \theta_{\rm weak}}~~. \eqno (82)$$
The global Techni-isospin group $SU_V (2)$ is not spontaneously
broken.  It is a manifest symmetry that implies, for example, $f_{T
\pi^{\pm}} = f_{T \pi^0}$.  Thus the ratio $(m_{W^{\pm}}/m_Z^0)$
is the same as
in the S.M.  Thus $\rho = 1$ is a consequence of a residual global
symmetry .
  This is a very elegant way to give masses to the $W$ and $Z$.
\b
Problems arise when trying to give masses to the standard fermions.
 In the S.M., these fermions receive masses through the
Yukawa couplings to the Higgs field, Eq. (62),
$$ {\cal{L}}_y = \lambda {\bar{f}}_L f_R \phi~~, \eqno(83)$$
when $< \phi > = v \not=0.$  Here we have $< \bar{Q} Q > \not=0$.  Thus
the obvious idea is to replace $\phi$ by $ \bar{Q} Q$ in Eq. (83),
$$ {\cal{L}}_{\rm mass} = \lambda {\bar{f}}_L f_R {\bar{Q}}_L Q_R ~~.
\eqno(84)$$
This is a non-renormalizable four-fermion interaction, of the type
encountered in weak interactions at scales $q < < m_W$.  We
envisage the same situation here, and interpret
Eq. (84) as the effective interaction
resulting at low energies from the exchange of a heavy gauge boson, with
mass $M_{ETC}$.  These bosons  mediate interactions between
fermions and Technifermions. Since they are massive, they must be the gauge
bosons of a spontaneously broken gauge interaction, called Extended
Technicolour (E.T.C.) $^{24}$.
 Consequently, from Eq. (84) we get
$$ m_f \simeq {< \bar{Q} Q >\over M^2_{ETC}}~, \eqno(85)$$
where the factor  $M^2_{ETC}$ comes from a bosonic propagator at low energies,
as can be seen by  rewriting Eq. (84) using a Fierz-transform as
$$ {\cal{L}}_{\rm mass} = ({\bar{f}}\gamma_{\mu} Q) {1\over M^2_{ETC}}
(\bar{Q} \gamma^{\mu} f)~. \eqno(86)$$
Thus, we have to assume a larger gauge group $G_{ETC}$ somehow broken
spontaneously into the group in Eq.(79).  In most E.T.C. models, the E.T.C.
generators commute with the S.M. ones, i. e. the E.T.C. gauge bosons are
$SU_C(3) \times SU_L(2) \times U_Y (1)$  singlets. In this case
 we need at least one full generation of
Technifermions, each one giving mass to the corresponding fermion with
the same flavour.  If
the E.T.C. generators do not commute
with the S.M. generators, the E.T.C. group and the S.M. group
should be unified into a larger group with a unique coupling constant.
However, since the couplings of the S.M. are very small at $500~GeV$,
while $\alpha_{ETC} \simeq O(1)$, one can unify them only by
introducing hugely different mass scales, a new gauge hierarchy problem
as bad as the one we wanted to avoid in the first place.

 The E.T.C. interactions
connect fermions from the three light generations, with
Technifermions, namely, light fermions and Technifermions are in
 one multiplet
of the E.T.C. group. Thus E.T.C. interactions
also produced flavour changing neutral currents
(FCNC) among the standard fermions, which yield important lower
 bounds on the
E.T.C. gauge boson masses.

The most restrictive bounds are those on FCNC in $\Delta s=2$
operators, such as
$$ {1\over M^2_{ETC}}~\bar{s}~d~\bar{s} d  \eqno(87)$$
which contribute to the $K_L - K_S$ mass difference, impose$^{25}$
$$ M_{ETC} > 500~ TeV~.  \eqno(88)$$

If the $G_{ETC}$ is broken in one stage, there is only one mass scale
$M_{ETC}$.
Just using dimensional arguments we obtain a naive estimate for $<
\bar{Q} Q >$, Eq. (76), namely
$$ < \bar{Q} Q > ~\simeq~ \Lambda^3_{TC}~. \eqno(89)$$
 Eqs. (85), (88) and (89), imply the upper bound
$m_f < 0.5~MeV$, too low for most fermion
masses.

 Recent attempts to raise the upper bound on $m_f$ are based on
obtaining  values larger than Eq. (89) for $< \bar{Q} Q>$, of the form
$$ < \bar{Q} Q > ~\simeq~ \Lambda^3_{TC} \left({M_{ETC}\over
\Lambda_{TC}}\right)^{\gamma}~, \eqno(90)$$
with $\gamma > 0$. In
asymptotically free theories the running coupling constant
 approaches zero as
the momentum transfer $q^2 \to \infty,~ \alpha_s \to 0$
($\alpha_s = 0$ is an
ultraviolet fixed point). In ``walking" or ``stagnant" T.C.
models $^{25,~26,~27}$, the rate at which
the running coupling constant approaches zero at large $q^2$
is decreased.  If
the evolution stops all together, one says that the coupling constant
``stands".  When it decreases with $q^2$ but not as fast as a
running coupling,
one says that the coupling constant ``walks".  A larger value
of the coupling constant $\alpha_{TC}$ as $q^2$ increases leads to the
formation of larger condensates.  In order to alter considerably
 the running of
$\alpha_s$ towards zero one needs to either introduce a
large number $n_f > >
2$ of Techniquarks $Q$ in the fundamental representation
of $SU(N_{TC})$, or
take just two flavours $n_f = 2$ of Techniquarks $Q$ but
 in a very large
representation of $SU(N_{TC})$.  In these models one
obtains $\gamma \leq~ 1$, which
yields a bound $m_f < 0.5~ GeV$,  still too low to
account for the $c, b$, and $t$ quarks.

To obtain larger values of $\gamma$ one needs a ``strong" E.T.C.
interaction$^{27,~28}$, i.e., not only $\alpha_{TC} \simeq  O(1)$ is
required, but also $\alpha_{ETC} \simeq O(1)$, so that E.T.C.
interactions help the T.C. interactions in the formation of the $<
\bar{Q} Q >$ condensates.  Here a fine tuning of $\alpha_{ETC}$ is
required, so that E.T.C. does not replace T.C. altogether (otherwise one
obtains $m_{W^{\pm}, Z} \simeq M_{ETC}!$).  In ``strong" E.T.C. models, one
obtains $\gamma \leq 2$, which yields the acceptable bound
$m_f < 500 GeV$.  Still, there are
usually some additional problems in these models, such as $\rho \not=
1$,
or an incorrect $m_t/m_b$ ratio, and the prediction of light scalar
bound states.

If the E.T.C. group is broken in several
stages at different scales $M_1 > M_2 > M_3 $ etc.,
$$ (G_{ETC})_1 \buildrel{M_1}\over{\longrightarrow}
 (G_{ETC})_2 \buildrel{M_2}\over{\longrightarrow}
 (G_{ETC})_3  \buildrel{M_3}\over{\longrightarrow}  \cdots
\eqno(91)$$
the hierarchy of E.T.C. gauge boson masses can be used to
arrange for the
hierarchy of fermion masses. In Eq. (85) the heavier E.T.C.
 bosons could be
used to give masses to the lighter fermions, and the lighter
 E.T.C. bosons to
give mass to the heavier fermions,for example,  $m_t~\simeq
\Lambda^3_{TC}/M^2_{\rm lightest}$.  In order to avoid FCNC
when the bound in
Eq. (88) is violated, one must postulate that E.T.C.
interactions couple the
fermions of different generations to different Technifermions
(so, for example
$u$ couples only to $Q_1$, $c$ to $Q_2$, $t$ to $Q_3$, etc.).
  Thus one needs
a T.C. model with at least three Technigenerations.

 Precision tests at LEP constrain T.C. models,  by its radiative
  effects in the
propagator of the electroweak gauge bosons$^{29}$. Models
with QCD-like
weak-isospin preserving T.C. dynamics with  one  whole family
of Technifermions
have been ruled out for all T.C. groups, $(N_{TC} \geq 2)$.
 Only a smaller
number of Technifermions, may be a single doublet of colour singlet
Techniquarks $U, D$, are still allowed$^{29}$ in models of this type.
Models with non-QCD- like dynamics, such as ``walking" T.C. and
``strong E.T.C." models,  may allowed$^{30}$.
 In particular, one- Technifermion- family models   with weak
 isospin splitting and  Techniquark-Technilepton splitting due
  to QCD and E.T.C.
interactions, may be allowed$^{31}$.
However, it  is difficult to reliably estimate the radiative effects
in such models, precisely because one cannot use analogies with QCD$^{32}$.

 So, there are still hopes for T.C., but no realistic theory is in sight.
Anyhow, T.C. predicts many bound states of Technifermions at scales above
$\Lambda_{TC}~\simeq~ 0.5~TeV$ that should be seen at SSC/LHC energies
and many light ones.  For example, a Technirho $T\rho$ should have a
large coupling to a pair of Technipions $ T \pi,~ T \pi~ \to T
\rho$.  The $T \pi$ are ``eaten up" by the $W$ and the $Z$, thus $T
\rho$ should appear as a resonance in $WW, WZ {\rm or} ZZ$ scattering.
  Other pseudo- Nambu- Goldstone bosons may be very light,
   $m \simeq O(100)~GeV$ or lighter.  In fact,
unsuccessful searches for these light bosons at LEP constrain
 many T.C. models.
\b
\no {\bf 5. ~Top quark condensate}
\b
  Because the top quark mass is almost as large as
 the Technifermion masses, the E.T.C. bosons in ``strong E.T.C."
  models
couple strongly not only to Technifermions but also to
$t$ quarks, and may generate $\bar{t} t$ condensates,
in the  same way they originate $\bar{F} F$ condensates.
 Having  this in mind, we could
ask ourselves if it would be possible to eliminate the
Technifermions and have the dynamical
breaking only through $< \bar{t} t> \not= 0$. The answer seems to be,
yes.  Top quark condensate models$^{33, 34}$ assume that some heavy
bosons of mass $\Lambda$, exchanged between $t$-quarks, generate, at low
energies an effective four fermion interaction
$$ {\lambda\over \Lambda^2} ({\bar{Q}}_L t_R) ({\bar{t}}_R Q_L)~~.
\eqno(92)$$

At scales $\mu < < \Lambda,~~ \phi = ({\bar{t}}_R Q_L)/\Lambda^2$ becomes
a composite Higgs field with V.E.V. $< \phi > = < \bar{t}_R Q_L
>/\Lambda^2 \not=0$ and a mass $M_H$ not very different from the
$t$-quark mass. The S.M. is obtained as a low energy effective
theory at $\mu < < \Lambda~~^{34}$,
$$\eqalign{{\cal{L}}_{\rm eff} = \lambda_t {\bar{Q}}_L t_R \phi + h.c. +
\cdots + \vert D_{\mu} \phi \vert^2
+ m^2 \phi^+ \phi - {\lambda\over 2} (\phi^+ \phi)^2 + O ({1\over
\Lambda^2})\cr} \eqno(93)$$
but with the compositeness boundary conditions
$$\eqalign{ \lambda_t (\mu), \lambda (\mu) & \bigg\vert \to \infty \cr
& \mu \to \Lambda \cr} \eqno(94)$$
that show that new physics comes into play at $\mu\simeq \Lambda$.
For $\Lambda = 10^{17}~GeV~~^{34}$, one
obtains $m_t = 220~GeV$, $m_H = 250~GeV$.  Such a high value
 of $\Lambda$ implies fine tuning
and no possibility of experimentally testing the compositeness scale.
For low values of $\Lambda$, that could be tested experimentally, one
gets much larger $t$-masses.  For example, $m_t = 450~GeV$ and $m_H =
600~GeV$, for $\Lambda = 10~TeV~~^{34}$.  In any case the values of $m_t$
obtained exceed present upper bounds.

These predictions have been criticized$^{35}$  because other higher
dimensional terms (besides $\bar{Q} t \bar{t} Q)$ may appear in the
non-renormalizable Lagrangian, Eq.(92).  These terms may alter the
predictions so much that predictivity is lost.

Also the idea of condensates of the quarks $(t^{\prime}, b^{\prime}) $
of a 4th. family has been conside- \break red$^{34}$.
Some of the results of these models are (for $m_t^{\prime} =
m_b^{\prime})$: $m_t^{\prime} = 200~ GeV$ and $m_H = ~240~ GeV$
 for $\Lambda
= 10^{17}~GeV,~ m_t = 400~ GeV$ and $m_H = 400~ GeV$ for $\Lambda = 10~ TeV$,
and $m_t = 1~ TeV$ and $m_H = 2~ TeV$ for $\Lambda = 2~ TeV$. The
main predictions are:  no light Higgs field and a 4th. family of
quarks that should be seen at LHC/SSC.  There are again some problems in
model building:  complicated models to avoid FCNC and the need for
four-fermion operators again.
\b
\no {\bf 6. ~Supersymmetry}
\b
Supersymmetry~$^{11}$ ensures that there are bosons and fermions,
supersymmetric partners of each other, that differ only in the spin.  A
supersymmetric generator $Q$ has spin $1/2$, since applied to a boson
state $\vert b >$ gives a fermion $\vert f>$ and vice versa,
$$ Q \vert b > = \vert f >~,~~ Q \vert f > = \vert b >~. \eqno(95)$$
In $N = 1$ supersymmetry there is only one generator $Q$, thus a
fermion-boson pair
$(b,f)$ form a supersymmetric multiplet.  $Q$, that is a fermion,
 satisfies the anti-commutation relations
$$ \{ Q, \bar{Q} \} = 2 \gamma^{\mu} P_{\mu}~, \eqno(96)$$
where $P_{\mu}$ is the generator of space-time translations, $P_0 =
H$, $P_i = {\vec{p}}_i$.  Eq. (96) tells us that
the anti-commutator of two global supersymmetric transformations constitute a
translation in space time, identical at any space-time point because
the transformations are global.  If the two supersymmetric
transformations are local, i.e. space-time dependent, its
anti-commutator gives a local translation.  A local translation in a
fixed reference frame, can also be understood as a local coordinate
transformation at a fixed point (either an observer is moving in a
fixed reference frame,   or the observer is at rest but the frame moves).
Thus, invariance under local supersymmetry implies invariance under
general space-time coordinate transformation (i.e. general relativity).
Since, local supersymmetry incorporates gravity, it is called
supergravity$^{16}$.

In $N= 1$ supersymmetry there is one ``sparticle" for every particle.
The minimal number of independent degrees of freedom in a fermion (a
Weyl fermion) is two.  Thus the minimal supersymmetric multiplet
involves a complex scalar field and  Weyl spinor.  Examples are:  $(\nu_L,
{\tilde{\nu}}_L)$, i.e. left-handed neutrino and left-sneutrino, $(e_L,
{\tilde{e}}_L)$ and $(e_R, {\tilde{e}}_R)$, i.e. electron and
selectron.  ``Left" and ``right" indices applied to scalars
 indicate that their fermionic partners
are left or right-handed.  The fermionic partners of bosons are called
``...ino".  For example, ${\tilde{\gamma}}, {\tilde{W}}, {\tilde{Z}},
{\tilde{G}}, {\tilde{H}}$, are called photino, W-ino, Zedino,
gluino and Higgsino.

Supersymmetry is not a good symmetry, since we know of no
degenerate boson-fermion pair. As mentioned earlier, Eq. (71), if
supersymmetry is to explain the stability of the weak scale against
radiative corrections the mass difference between supersymmetric partners
should be
$$ m_{\rm sparticle}~ -~ m_{\rm particle}~ \leq~ O (1~ TeV)~. \eqno(97)$$

There is another rule that early model builders discovered: no known
particle can be the sparticle of any other.  Thus, the sparticles must
all be new particles (see Table 2.))
\bb
\no Table 2. Spin of the sparticles corresponding in $N=1$ supersymmetry
 to the known particles.
\input tables.tex
\thinsize=.25pt
\thicksize=.25pt
\begintable
\multispan{2}\tstrut\hfill{\bf Particle}\hfill | \multispan{2}\tstrut\hfill
{\bf Sparticle}\hfill \crthick
$S=0$ & Higgs (H) | $S=1/2$ & Higgsino ($\tilde{H}$)\crnorule
$S=1/2$ & fermion | $S=0$ & sfermion\crnorule
& $(\ell_{L,R}~,\quad q_{L,R})$ | &  $(\tilde{\ell}_{L,R}~,\quad
\tilde{q}_{L,R})$\crnorule
$S=1$ & gauge bosons | $S=1/2$ & bosino\crnorule
& $(\gamma,\th Z,\th W,\th gluon)$ | &  $(\tilde{\gamma},\th \tilde{Z},\th
\tilde{W},\th \tilde{g})$\crnorule
$S=2$ & graviton | $S=3/2$ & gravitino\endtable
\b
Supersymmetry and gauge invariance insure that for every vertex of
particles, there must be the corresponding ones with sparticles.  So,
for example, to
$\gamma \to e^+ e^-$ correspond $\tilde{\gamma} \to {\tilde{e}}^+
e^-, e^+ {\tilde{e}}^-$ and $\gamma \to {\tilde{e}}^+
{\tilde{e}}^-$.  The conservation of lepton and baryon numbers, L and B,
usually require discrete symmetries, the most common of which is
$R$-parity$^{36}$. It is defined for each particle or sparticle as
$$ R = (-1)^{3(B-L) + 2S} \eqno (98)$$
what means $ R = +1$ for a particle and $R = -1$ for a sparticle.
$R$-conservation has two very important consequences:
\s
\item{-} sparticles are produced in pairs
\item{-} the lightest sparticle
(the LSP, lightest supersymmetric particle), is stable;
usually it is a neutral particle that escapes detection
 in collider experiments.
\s
\no Thus, while $R$-parity is conserved the signature
 for supersymmetry in colliders is, precisely,
supersymmetric particles produced in pairs that end-up producing missing
energy and momentum.  One should keep in mind that $R$-parity could be
slightly violated, either explicitly$^{37}$, or spontaneously through a
non zero V.E.V. of a sneutrino, $< {\tilde{\nu}}_{\tau} > \not=0^{38}$.
This violation would have interesting experimental consequences$^{39}$.

 Supersymmetry allows for the coupling of three multiplets such as the
 Yukawa coupling $QD^cH$, for example, but not of multiplets and
complex conjugates of multiples, such as $QU^cH_1^c$ (this is
\underbar{not} allowed).  Thus, we cannot use just one Higgs field
to give mass to all fermions, as we do in the S.M. (where we use $\phi$
for the up-fermions and its conjugate $\tilde{\phi}$ for the down-fermions,
 Eq.'s (58) and (62)).  We need two
Higgs doublets $H_1$ and $H_2$, to give mass to the down-fermions and to
the up-fermions, respectively.  Anomaly cancellation also requires two
Higgs multiplets.
There are eight real  components in two complex doublet Higgs
fields.  Through the Higgs mechanism three of them are ``eaten up" by
the $ W^{\pm}$ and $Z^0$.  Five remain as physical particles:  $H^{\pm},
h^0, H^0, A^0$.  Thus, there are more Higgses than in the S.M.
to look for, two charged
ones and three neutral ones.

After many attempts to build phenomenologically viable supersymmetric
models, those with supergravity spontaneously broken at a large energy
scale, close to the Planck mass $M_P$, have been proven to be the best.
Supergravity is broken by fields in a ``hidden" sector of the theory,
i.e. by fields coupled only gravitationally to the standard sector.  The
 spontaneous breaking of supergravity produces a non zero mass for
the gravitino, $m_{3/2} \not= 0$, and/or gauginos, $m_{1/2} \not= 0$.
At low energy scales (in the limit $M_P \to \infty$, with $m_{3/2}$ and
$m_{1/2}$ fixed), what remains is a theory with supersymmetry explicitly
broken by ``soft" breaking terms, i.e. terms with mass dimension 2 and
3.  Since the mass dimension of a bosonic field, $b$, is 1 and that of a
fermion  field, $f$, is 3/2, the terms with dimension 2 and 3 are of
the form
$$ b^{\dagger} b,~b^2, ~b^3, \bar{f} f~. \eqno(99)$$
Supersymmetry is broken by all possible gauge invariant terms
of these types.  From terms $\tilde{q}^{\dagger}~ \tilde{q}$ and
$\tilde{\ell}^{\dagger}~\tilde{\ell}$,
we get large
masses for the $\tilde{q}$ and
$\tilde{\ell}$,  $m_{\tilde{q}}~ >  m_q$, $ m_{\tilde{\ell}} >
m_{\ell}$;  from terms of
the form $ \tilde{\gamma} \tilde{\gamma}$, etc.,we get large
 masses for all the gauginos,~
$\tilde{\gamma}, {\tilde{W}}^{\pm}, \tilde{Z}, \tilde{g}$; besides,
we get all other possible
quadratic and cubic couplings among scalars.  Notice the amazing property
of these models:  the explicit
supersymmetry breaking terms  provide large masses
only for the sparticles.
While the gauge symmetry is unbroken,
the masses of the standard particles, quark, leptons and
gauge bosons, are zero. These particles get  masses,
due to the
Higgs mechanism, only when the Higgs fields develop
non-zero V.E.V.'s.
 Thus, these models
have to incorporate a mechanism to get non-zero V.E.V.'s for
the Higgs fields, $< H_i >= v_i \not= 0$, and not for other
 scalars, such as
$\tilde{q}$ and $\tilde{\ell}$, $< \tilde{q} > = < \tilde{\ell} > = 0$.
\b
\no {\it 6.1. ~The Minimal Supersymmetric Standard Model (M.S.S.M)}
\s
The M.S.S.M.$^{40,~41}$ is the simplest phenomenologically acceptable
supersymmetric version of the S.M. The model has $N = 1$ supersymmetry
spontaneously broken at large energies $ \simeq
M_P$ by a ``hidden"
sector,  and supersymmetry broken by ``soft" terms, as explained above.
The M.S.S.M. has several additional defining properties$^{40}$.

 \no 1)  At the
TeV energy scale the symmetry group is that of the standard model,
$SU(3) \times SU(2) \times U(1)$ (models with additional symmetries,
such as an extra $U(1)$ for example, are non minimal).

\no 2) There are
no new matter fields besides the standard quarks and leptons and their
corresponding supersymmetric partners, the sleptons and squarks.

\no 3) The Higgs sector contains just two doublets $H_1$, and $H_2$,
 the first
coupled to the down fermions and the second coupled to the up ones
(therefore, after the electroweak symmetry breaking,
there are three neutral and two charged physical Higgs bosons
in the model; models with more Higgs fields are non-minimal).

\no 4) Simplifying assumptions leave only
five parameters in the Lagrangian, at the scale of Grand Unification, in
addition to those of the S.M.   The Lagrangian
$$ {\cal{L}} = {\cal{L}}_{\rm SUSY} + {\cal{L}}_{ \rm SUSY~ breaking}
\eqno(100)$$
has a supersymmetry breaking part
$$\eqalign{ {\cal{L}}_{\rm SUSY-breaking}& = \sum_i m_{0_{i}}^2 \vert
\phi_i \vert^2 + \sum_{\alpha} M_{\alpha} \psi_{\alpha} \psi_{\alpha} + \cr
& + \sum_{\lambda} A_{\lambda} \lambda \phi^3 + B \mu H_1 H_2 +
\sum_{\delta} B_{\delta} \delta \phi^2 + h.c. \cr} \eqno(101)$$
where $i$ labels all scalars, $\alpha = 1,2,3$, labels the three
gauginos, $\lambda$ labels all cubic coupling constants, $\mu$ is a
supersymmetric mass term for the Higgs fields and $\delta$ labels other
quadratic coupling constants.  The constants $m_{0_{i}}, M_{\alpha},
A_{\lambda}, B, B_{\delta}$ depend on the ``hidden sector".  In the
M.S.S.M. it is assumed that at a large scale $M_U$ (that could be a Grand
Unification scale):
$$ m_{0_{i}}\bigg\vert_{M_U}\! \! \! \!\equiv m_0~~,~~ M_1
\bigg\vert_{M_{U}}\! \! \! \! = M_2 \bigg\vert_{M_{U}}\! \!
\! \! = M_3 \bigg\vert_{M_{U}}\! \! \! \! \equiv m_{1/2},~~ A_{\lambda}
\bigg\vert_{M_{U}}\! \! \! \! \equiv A,~~B_{\delta} \bigg\vert_{M_{U}}
\! \! \! \! \equiv B~~. \eqno(102)$$
Thus, The M.S.S.M. has five parameters (besides those of the S.M.).  They are
$m_0$, a common mass for the scalars, $m_{1/2}$, a common mass for the
gauginos, $\mu$, a supersymmetric
mass term for the Higgs fields, and two adimentional parameters, A and
B.  Many times there is a further reduction of the number of parameters.
Two of the most used choices are either $ B = A  - 1$, an older choice,
or $m^2_{1/2} > > m^2_0, A^2~ \simeq~ 0$ and $B^2~\simeq~ 0$, a ``string
inspired" choice$^{42}$.
\b
The theory has to be scaled down, from the GUT scale to low energies,
according to the renormalization group equations.  Through radiative
corrections, the Higgs masses become negative. Thus, the Higgs fields
develop non-zero vacuum expectation values, $< H_1 > = v_1$ and $< H_2 > =
v_2$, while the other scalar fields do not, $< \tilde{q} >
= < \tilde{\ell} > = 0$. This is a nice feature of
the M.S.S.M. This property is due to the effect of the heavy top quark,
that couples to $H_2$, driving its mass negative.  A correct spontaneous
breaking of the standard model implies two minimization conditions, that
reduce the five parameters mentioned above to just three independent
parameters.  They can be chosen to be a gaugino mass,
 for example $M_2$, the Higgs
mass parameters $\mu$, and the mass of the lightest scalar Higgs field,
$m_h$.  $M_1, M_2$ and $M_3$ are the masses of the three gauginos, the
supersymmetric partners of the $U_Y(1), SU_L(2)$ and $SU_C(3)$ gauge bosons
respectively.  This masses, assumed in the M.S.S.M to be all
identical to $m_{1/2}$ at the scale $M_U$ (Eq. 102) get renormalized to
$$ M_3 = (\alpha_s/\alpha) \sin^2 \theta_W~ M_2 = ( 3 \alpha_s/5 \alpha)
\cos^2 \theta_W~ M_1~, \eqno(103)$$
i.e. $M_3~\simeq~ 4 M_2~\simeq~ 8 M_1$, at the weak scale $(\alpha_s$ is
the strong coupling constant, $\alpha$ the electromagnetic coupling
constant, and $\theta_W$ is the weak mixing angle).

It is the assumption of equal gaugino masses at a unification scale,
that
allows  translating the CDF lower bound on the gluino mass $M_3$, into a
lower bound of $10 - 20$~GeV for the lightest neutralino mass, since
this mass depends on $M_1$, and $M_2$.  Not making this assumption leads
to a non-minimal model$^{43}$ where neutralinos may be lighter than
$10~GeV$.
\b
The M.S.S.M. incorporates the simplest viable assumptions, however, one
or several of them may be proven wrong.  We can keep the M.S.S.M. as
long as it works, and we do not have better motivations to study more
complex models.
Some possible departures from the M.S.S.M.
have been explored:  models with broken R-parity, i.e. with L or B
violations (so the lightest sparticle is not stable and sparticles are
not produced in pairs)$^{37,~38,~39}$, models with extended
 Higgs sectors (for example,
with one extra singlet$^{44}$), models with non universal supersymmetric
breaking terms (in which the assumptions in Eq.(102) are modified$^{43}$).
\b
{\it 6.2. ~Charginos and Neutralinos in the M.S.S.M.}
\s
In the M.S.S.M. there are four ``neutralinos", two neutral gauginos
${\tilde{W}}_3$ and ${\tilde{B}}$ (the supersymmetric partners of the
3rd. $SU(2)$ gauge boson, $W^3_{\mu}$, and the $U(1)$ gauge boson,
$B_{\mu}$, see Eq.(34) and two neutral Higgsinos, ${\tilde{H}}_1^0$
and ${\tilde{H}}_2^0$ (the partners of the neutral components of the two
Higgs doublets).  In the base $({\tilde{B}}, {\tilde{W}}_3,
{\tilde{H}}_1^0, {\tilde{H}}^0_2)$, their four by four mass matrix is:
$$ \pmatrix{ M_1 & 0 & -M_Z c_{\beta} s_W & M_Z s_{\beta} s_W
 \cr
 0 & M_2 & M_Zc_{\beta} c_W & -M_Z s_{\beta} c_W \cr
-M_Z c_{\beta} s_W & M_Z c_{\beta} c_W & 0 & - \mu \cr
M_Z s_{\beta} s_W & -M_Z s_{\beta} c_W & -\mu & 0 \cr}~~, \eqno(104)$$
with the abbreviations $c_{\beta}, s_{\beta}, c_W$ and $s_W$ for $\cos
\beta, \sin \beta, \cos \theta_W$ and $\sin \theta_W$, respectively.
Here $\sin^2 \theta_W = 0.23$ and $\tan \beta \equiv v_2/v_1$.  $M_1$ is
related to $M_2$ through Eq.(103).  There are four mass eigenstates
${\tilde{\chi}}^0_i$, ($i = 1, 2, 3, 4$, in order of increasing mass),
that are in general a linear
combination of different current eigenstates (photino, Zino and
Higgsino). In many models, the lightest neutralino is  the LSP, the lightest
supersymmetric particle, and one of the preferred D.M. candidates.  It is
in general
$$ {\tilde{\chi}}^0_1 = \gamma_1 \tilde{\gamma} + \gamma_2 \tilde{Z} +
\gamma_3 {\tilde{H}}^0_1 + \gamma_4 {\tilde{H}}^0_2~, \eqno(105)$$
where the coefficients
$\gamma_i$ depend only on $M_2, \mu$ and $v_2/v_1$.  When $M_2
\to 0$, the $ {\tilde{\chi}}^0_1$ is the photino $\tilde{\gamma}
 = \cos \theta_W
\tilde{B} + \sin \theta_W {\tilde{W}}_3$, with mass proportional to
$M_2$.  When $\mu \to 0$, the ${\tilde{\chi}}^0_1$ is
the Higgsino $\tilde{h} = \sin
\beta {\tilde{H}}^0_1 + \cos \beta {\tilde{H}}^0_2$, with mass
proportional to $\mu$.  A general ${\tilde{\chi}}_1^0$ as in Eq.(105) is
usually heavier than the pure $\tilde{\gamma}$ or $\tilde{h}$ (see Fig. 3.).
\vglue 9.truecm
\no Fig. 3. Contour map of the mass of the lightest neutralino in $GeV$
and its dominant component (in regions separated by dotted lines) in the
$(\mu - M_2)$ plane, for $\tan \beta = 2 $. $^{45}$
\bb

\no In this
model the mass of the lightest neutralino can be anything between zero
and a few TeV.  The upper bound is a matter of consistency with the
motivation to introduce supersymmetry  as a solution to the hierarchy
problem, Eq.(97). When ${\tilde{\chi}}^0_1$ is much heavier
 than the $Z$-boson,
it is a pure $\tilde{B}$ for $M_2 < \mu$, and a pure
Higgsino for $M_2 > \mu$. This can be easily seen by
 setting $M_Z = 0$ in the matrix in Eq.(104).
Usually, all parameters of the Lagrangian are taken to be real,
 neglecting possible
small CP violation effects.  Without loss of generality, $M_2$ is chosen
non-negative.  A correct electroweak breaking requires $v_2/v_1 > 0$.
For a heavy top, $m_t \geq~120$~GeV, a value as large as 20 for
this ratio may be  allowed.
\b
The ``chargino" masses depend on the same three parameters, $M_2, \mu$ and
$v_2/v_1$.  The charginos are the two Dirac fermion mass eigenstates.
 They are linear combinations of the ${\tilde{W}}^-$, the
${\tilde{W}}^+$, the ${\tilde{H}}_1$ and the ${\tilde{H}}^+_2$.  These
are the superpartners of the $W^{\pm}$ gauge bosons, and of the charged
components of the two Higgs fields.

The mass matrix of the charginos in the base $({\tilde{W}}^+,
{\tilde{H}}^+_2, {\tilde{W}}^-, {\tilde{H}}^-_1)$ is given by
$$ \pmatrix{ 0 & 0 & M_2 & M_W \sqrt{2}c_{\beta} \cr
0 & 0 & M_W \sqrt{2}s_{\beta} & \mu \cr
M_2 & M_W \sqrt{2}s_{\beta} & 0 & 0 \cr
M_W \sqrt{2}c_{\beta} & \mu & 0 & 0 \cr} ~~. \eqno(106)$$
There are two chargino mass eigenstates ${\tilde{\chi}}^{\pm}_i$, $i = 1,
2$.
\b
{\it 6.3. ~Sparticle Searches}
\s
At $e^+ e^-$ colliders, bounds have been obtained from
the non-observation of $Z$ boson decays
into sparticle pairs, ${\tilde{\ell}}^+ {\tilde{\ell}}^-$,
 $\tilde{\nu} \tilde{\nu}$, $\tilde{q} {\buildrel\simeq\over
q}$,
${\tilde{\chi}}^+ {\tilde{\chi}}^-$, ${\tilde{\chi}}^0_i
{\tilde{\chi}}^0_j $.  Both, direct searches and indirect limits from
partial decay widths, give roughly a bound $\tilde{m} > m_Z/2~$ for all
sparticles, with two exceptions:  the lightest neutralino,
 ${\tilde{\chi}}^0_1$, and a light stop, ${\tilde{t}}$, because mixing
effects can decouple them from the $Z$.  Usually the squark masses are
given by $m_{{\tilde{q}}_{L,R}} = \vert m_0 \pm m_q \vert$,
with $m_0$ a
common scalar mass (see Eq. (102)).  Because  $m_t$ is large,
$m_{\tilde{t}} = m_0 - m_t$ can be small.  The lower bound on
$m_{\tilde{t}}$ is 20 GeV$^{46,~41}$. The lightest neutralino mass is larger
than $20~GeV$ as soon as $v_2/v_1>3$, $^{47}$ and bounds on
 charginos and neutralinos can be seen in Fig. 4. (taken from Ref. 48).
It is interesting to point out that the M.S.S.M. is as good as the S.M. with
respect to bounds coming from precision measurements at LEP$^{49}$.

The bounds from $ p \bar{p}$ colliders are better for strongly
interacting sparticles.  These particles are produced in pairs $\tilde{g}
\tilde{g},~ \tilde{q} \tilde{q},~ \tilde{q} \tilde{g}$.  Their production
rate is relatively model independent.  The signatures depend on the
relative masses.  If $m_{\tilde{g}} < m_{\tilde{q}}$,
then squarks decay first,
$\tilde{q} \to q \tilde{g}$, and one searches for gluino
decays, $\tilde{g} \to \bar{q} q{\tilde{\chi}}^0_i,~
\bar {q} q^{\prime}
{\tilde{\chi}}^+_j$.  If $m_{\tilde{q}} < m_{\tilde{g}}$
instead, gluinos
decay first, $\tilde{g} \to \tilde{q} \bar{q}$, and one
searches for
squark decays $\tilde{q} \to q {\tilde{\chi}}^0_i,~
q^{\prime}
{\tilde{\chi}}_j^+$.  In both cases it is important to
include
``cascade" decays$^{50}$, in which heavier neutralinos
and charginos can
be produced and decay into lighter ones.  The inclusion of ``cascade"
decays has recently led to a revision of the CDF bounds on
$m_{\tilde{q}}$
and $m_{\tilde{g}}~~,^{51}$. The signatures are
missing energy and equal sign dileptons or diquarks.  Equal sign
difermions are due to the Majorana nature of $\tilde{g} $
and \x, that
are their own antiparticles.  Thus, the same decay products are
possible on both sides of the produced pairs.
\vglue 10.truecm
\no Fig. 4.  Contour map of the mass of the lightest neutralino
(dotted lines) and chargino (dashed lines) in $GeV$, showing
present limits (shaded area) and future sensitivity of the complete
LEP and Tevatron programs (solid lines),  in the
$(\mu - M_3)$ plane (where $M_3$ = $m_{\tilde{g}}$),
for $\tan \beta = 2 $. $^{48}$
\bb
 The lightest neutralino as the LSP, either light or
 heavy as compared with the
weak gauge bosons,  is one of the theoretically
better motivated and testable
D.M. candidates.  Unless a very odd origin of the
neutralino relic density is
invoked, such as a late decay of gravitinos, for
 example, the relic density of
neutralinos is determined by the same parameters
$(M_2, \mu, v_2/v_1, m_h,
etc)^{45, ~52,~53}$, that determine the experimental bounds.
 In  Fig 5. (taken from
Ref. 53) we  see in
gray the regions of the $(\mu, M_2)$ parameter space for
 which the M.S.S.M.
 lightest neutralino with $m_{{\tilde{\chi}}^0_i} < m_W$
could account for the D.M., having a relic density
 $0.25 < \Omega
h^2 < 0.5~^{53}$. The regions rejected by LEP and CDF
 are also shown.  As
explained above, the CDF bound on gluinos can be
translated into a bound
on \x,
because of the assumption of equal gaugino masses
at a high energy scale $M_U$.
\vfill \eject
\vglue 11.truecm
\no Fig. 5.  Contour of relic LSP neutralino density in the
$(\mu - M_2)$ plane, for $\tan \beta = 2 $ and the
indicated choices of other parameters. The calculation
includes radiative corrections to the Higgs masses
(see below).
For neutralinos with $m_{{\tilde{\chi}}^0_i} < m_W$
this figure  shows regions with  $0.25 < \Omega
h^2 < 0.5$ (shaded), where the neutralino is a good D.M.
candidate, regions excluded because
 $ \Omega h^2 > 1$  (coss-hatched) and regions excluded by
  LEP and CDF.
The $ \Omega h^2  = 0.1$ and $ \Omega h^2 = 0.025$ contours
 are shown as a thick dashed and  solid lines respectively $^{53}$.
\bb
 For neutralinos much heavier than the gauge bosons,
 only the regions with
$M_2~\simeq~ \mu~$ for which \x~ is a mixed gaugino -
 Higgsino state yield
densities suitable for a D.M. candidate$^{54}$.
For $M_2 > \mu$, where the
lightest neutralino \x is a pure Higgsino, the
density is too low ,
 $\Omega h^2
< 0.001$, because of the fast Higgsino annihilation $\tilde{H}
\tilde{H} \to
W^+ W^-$, in the early universe.  For $M_2 < \mu$, where \x =
$\tilde{B}$, a
bino, the relic density is too large $\Omega h^2 > 1$, because
 the dominant
annihilation mode, $\tilde{B} \tilde{B} \to H H$, has a very
small cross
section. Thus, bounds on the present total energy density of
 the universe
exclude this region.

\ b
\no {\it 6.4. ~Higgses and Top in the M.S.S.M.}
\s
As we mentioned above there are five physical Higgses (besides those
``eaten up" by gauge bosons), two charged ones, $H^{\pm}$, and
three neutral ones,
$h$ and $H^0$ that are CP-even fields, and $A$, that is CP-odd.  At
tree level, all the Higgs masses depend on only two parameters,
 usually
chosen to be $m_A$ and $\tan \beta = v_2/v_1$, and one can
prove the
following relationships
$$ m^2_{H^{\pm}} = m^2_W + m^2_A \eqno(107)$$
$$ m^2_h + m^2_{H^0} = m^2_A + m^2_Z \eqno(108)$$
$$ m_W, m_A < m_{H^{\pm}}~,~~~ m_h < m_A < m_{H^0}~.\eqno(109)$$
Combining Eqs. (109) and (108), we obtain $m_{H^0} > m_Z$ and
$m_h < m_Z$.
That is, the tree level relations predicts that the lightest
neutral Higgs
field,  $h$, is lighter than the $Z$.  After radiative corrections,
 mainly due
to $t$ and $\tilde{t}$ loops$^{55}$, the upper bound on the lightest
Higgs increases to
$$ m_h < ~150~GeV~. \eqno(110)$$
Thus the prediction of a ``light" Higgs in the M.S.S.M.
 persists $^{56,~41}$
(see Fig.6, taken from Ref. 56).

\vglue 10.5truecm
\no Fig. 6. Contours of the maximum value of  $m_h$,
 the lightest Higgs mass,
for $m_A >> m_Z$ and $\tan \beta >> 1$, in the
($m_t$, $m_{sq}$) plane$^{56}$.
\bb
The supersymmetric prediction of a light Higgs persists even in
non-minimal models.  For example, in the minimally
 non-minimal
supersymmetric standard model, M.N.M.S. S.M.$^{44}$
in which a singlet Higgs field is
added to the M.S.S.M. one obtains after including
radiative corrections$^{57}$
$$ m_h <~140~GeV~. \eqno(111)$$

With respect to the $t$ quark, typical values of
the unification scale,
$M_U~\simeq$ $10^{16}~GeV$ and of the sparticle masses
$\tilde{m}~\simeq~G_F^{-1/2}$ give an upper bound $^{41}$
$$ m_t~\leq~ 200~GeV~\sin \beta~. \eqno(112)$$
\b
{\bf 7. ~ Grand Unified Theories}
\b
GUT's attempt a ``true unification" of strong and
electroweak
interactions into a single simple non-Abelian
 gauge group G, with a
unique coupling constant $g$.  The GU symmetry
should break
down spontaneously into the S.M. symmetry in one
or several stages. We will
talk here about ``vertical" Grand Unifications that
do not attempt to provide
an explanation for the existence of more than one family.
Families still appear as
replicas - as they do in the S.M. ``Vertical" unification
 means ``intra-family"
unification.  The name stems from our habit of writing
families in columns.
\b
Within the S.M. there is no explanation for the values of the electrical
charges of quarks and leptons, since they depend on the arbitrary choice
of hypercharges $Y$.  In a GUT, we expect the choice of the values of
$Y$ to depend on other quantum number assignments in $G$.  Thus we
expect to explain why a proton and an electron have equal (and opposite)
charges to  at least one 1 part in $10^{21}$,
 i.e.~ $Q_u =$ $- (2/3)Q_e$ and $Q_d =$ $(1/3) Q_e$.

We also expect some other free parameters in the S.M. to become related
to each other, thus reducing the number of parameters.
\b
In GUT's, quarks and leptons are members of the same multiplet of $G$,
thus GU interactions can change one into the other, violating $B$ and
$L$.  A process such as $qq \to \bar{\ell} \bar{q}$, for example,
 mediated by a heavy
gauge boson $X$ of mass $M_X$, would induce proton decay,
$p^+ \to e^+ \pi^0$,  with a rate
$$ \Gamma~\simeq~ {g^4\over M_X^4}~m_p^5 ,\eqno(113)$$
where $m_p \simeq 1~GeV$ is the proton mass, and the factor
$m_p^5$ is just a reasonable guess to obtain the proper dimension
for $\Gamma$.  From this rough estimate and the experimental bound on
$p$-decay, $\tau_p \geq~ 10^{33}y$, $^{58}$ we obtain
$$ M_X ~\geq~ 10^{14}-10^{15}~GeV~.\eqno(114)$$

It is remarkable that
when the three running coupling constants of the S.M.
are normalized with a common convention, their logarithmic behavior
draws them together at approximately the same scale of Eq. (114).  In
fact, the running coupling constants of the S.M. at one loop, when
measured at specific values of momentum transfer $Q$ are given by
$$ \alpha_i^{-1} (Q^2) = \alpha_i^{-1} (\Lambda^2_i) + \beta_i ~\ell n
\bigl( {Q^2\over \Lambda_i^2}\bigr)~, \eqno(115)$$
where $\alpha_i = g^2_i/4 \pi~,~ i = 1,2,3$ refer to the $U_Y (1),~
SU_L (2)$ and $SU_c (3)$ couplings respectively, $\beta_3 = (33 - 4
N_G)/12 \pi,~~ \beta_2 = [(21/2) - 4 N_G]/12 \pi$, and
$\beta_1 = (-0.3 - 4 N_G)/ 12 \pi$
and $N_G $ is the number of generations.  Taking a common scale
$\Lambda_i = M_X$ where we assume $\alpha_1 = \alpha_2 = \alpha_3$,
and subtracting Eq. (115) for two of the constants,
we obtain $M_X/Q \simeq
O [exp (\alpha_i^{-1} - \alpha_j^{-1})]$. That provides
again the estimate in Eq. (114) for the ``unification scale" $M_X$.
\b
Since, as we mentioned above, the rank of the S.M. gauge group is 4,
the rank of the G.U. group G must be larger than
or equal to 4.  The
simplest rank-4 group that allows for a chiral embedding of the S.M.
(where right and left handed fields are treated differently), i.e. that
possesses complex representations, is $SU(5)$.
\b
{\it 7.1. ~The $SU(5)$ model $^{12,~4}$}
\s
The fundamental representation has dimension 5, and the $n^2 - 1 = 24$
generators $T^a$ can be written as $5 \times 5$ matrices.  These matrices
contain the eight $3 \times 3$ matrices $(\lambda^a/2)$, generators of
$SU_C(3)$, the three $2 \times 2$ matrices $(\tau^i/2)$, generators of
$SU_L(2)$ and a diagonal generator corresponding to $U_Y(1)$, in the
following way:
$$ \pmatrix{  {{\lambda^a} / {2}}~~~ &  0\cr
& \cr
0~~~ & 0 \cr}~,~~~~~~ I_i \equiv \pmatrix{ 0 ~~~~& 0 \cr
 & \cr 0~~~~ & {{\tau^i}/ {2}} \cr}~,~~~~~~
I_0 \equiv \pmatrix{ D_A~~ & 0\cr & \cr 0~~ & D_B \cr}~,
 \eqno(116)$$
where $a = 1, \cdots 8$ and $i = 1,2,3,~ D_A$ is a
$3 \times 3$ diagonal matrix
with elements $A,~ D_B$ is a $2 \times 2$ diagonal
 matrix with elements
$B$, and $A =
-2/\sqrt{60},~ B = -(3/2)A$ (notice that the condition
in Eq. (28), with $r = 1/2$, is satisfied) .  Besides
these generators, there are two sets of six matrices that have
non-zero elements in the $2 \times 3$ and
$3 \times 2$ horizontal and
vertical blocks not occupied in the matrices in Eq. (116), -
 which makes $8
+ 3 + 1 + 6 + 6 = 24$.  Notice that there are only four
diagonal matrices
(since the rank is 4), that correspond to $\lambda^3, \lambda^8, \tau^3$
and $I_0$. A gauge field multiplies each of the 24 generators
in the covariant derivative (Eq's (23) and (34)):
 gluons $G^a_{\mu}$ multiply
$\lambda^a/2$,
weak gauge bosons $W^i_{\mu}$ multiply $\tau^i/2,~
 B_{\mu}~{\rm to}~ I_0$, and
twelve new gauge bosons, the ``lepto-quarks"
$X_{\mu}, {\bar{X}}_{\mu}$ and
${Y}_{\mu},{\bar{Y}}_{\mu}$ multiply the 12
 generators in the non-diagonal
blocks.  The matrix ($\sum_b\th
{T^b} A^b_{\mu}$),  b= 1,2,$\cdots$,24,
in the $SU(5)$ covariant derivative (Eq. (23)),
that consequently appears in the couplings of the
 $SU(5)$ gauge bosons $A^a_{\mu}$
with fermions, is of the form
$$ \pmatrix{G_{\mu} + A B_{\mu} & G_{\mu} & G_{\mu} & {\bar{X}}_{\mu} &
{\bar{Y}}_{\mu} \cr
G_{\mu} & G_{\mu} + AB_{\mu} & G_{\mu} & {\bar{X}}_{\mu} & {\bar{Y}}_{\mu}
\cr
G_{\mu} & G_{\mu} & G_{\mu} + AB_{\mu} & {\bar{X}}_{\mu} &
{\bar{Y}}_{\mu} \cr
X_{\mu} & X_{\mu} & X_{\mu} & (W^3_{\mu}/2) + BB_{\mu} &
 (W^+_{\mu}/ \sqrt{2}) \cr
Y_{\mu} & Y_{\mu}& Y_{\mu} & (W^-_{\mu}/ \sqrt{2}) &
(-W^3_{\mu}/ 2) + BB_{\mu}
\cr}~~.\eqno(117)$$
\b
We use only left-handed spinors, $f_L$ and $(f^c)_L$, as independent
fields (as
explained following Eq. (20)).  The 15 Weyl spinors of each generation
are accommodated in the
representation $\bar{5} + 10$, without requiring any extra field,
$$ {\bar{5}}:\left(\matrix{~d^c_r \cr d^c_y \cr d^c_b \cr e \cr \nu_e
\cr}\right)_L= f_{{\bar{5}}}~~~~~~~ 10:{\pmatrix
{0 & u^c_b & -u^c_y &
-u_r & -d_r \cr -u^c_b & 0  & u^c_r & -u_y & -d_y \cr u^c_y &
-u^c_r & 0 & -u_b & -d_b \cr u_r & u_y & u_b & 0 & -e^c \cr d_r & d_y &
d_b & e^c & 0 \cr}}_L = f_{10} \eqno(118)$$
where $r, y, b$ stand for the three colours, red yellow and blue.
\b
 From Eqs. (117) and (118) we see that the lepto-quark gauge bosons are
those that couple leptons and quarks.  These are the bosons whose mass,
$M_X$, must fulfill the bound in Eq. (114).  This mass arises due to the
spontaneous breaking of $SU(5)$ into the S.M. group at the large scale
$M_X$.  This breaking is achieved with a Higgs boson $\Phi$ in the
representation 24.  This is the same representation of the gauge fields,
so we choose to write $\Phi$ as a $5 \times 5$ matrix.  The minimum of the
potential determines a non-zero V.E.V., $< \Phi >$, which we are free to
choose as a
diagonal matrix (through $SU(5)$ notations),

$$ < \Phi > = v_{\Phi} \pmatrix{ 1 & 0 & 0 & 0 & 0\cr 0 & 1 & 0 & 0 & 0 \cr
0 & 0 & 1 & 0 & 0 \cr 0 & 0 & 0 & -3/2 & 0 \cr 0 & 0 & 0 & 0 & -3/2
\cr}~~. \eqno(119)$$
This V.E.V. is left invariant by $SU_C(3) \times SU_L(2) \times U(1)$
transformation, i.e. it commutes with their generators given in Eq.
(116). But this V.E.V. is not invariant under any other transformations, i.e.,
it does not commute with the other 12 generators, which are, therefore,
spontaneously broken. Thus, we obtain masses for the $X$ and $Y$ leptoquarks.
$$ M_X^2 = M^2_Y = {25\over 8} g^2 v^2_{\Phi}~~. \eqno(120)$$
Thus, Eq. (114) implies $v_{\Phi} \geq 10^{14}~GeV$.  We have also to
introduce another Higgs field, H, in the representation 5, and arrange
its coupling to $\Phi$ so that when $\Phi$ has its V.E.V. at $< \Phi
>$,  Eq. (119), H has its V.E.V. at
$$ < H > = {v_H\over \sqrt{2}} \pmatrix{ 0 \cr 0 \cr 0 \cr 0 \cr 1
\cr}~. \eqno(121)$$
The lower two components of H act as the standard doublet Higgs field.
Thus $ < H >$ induces the electroweak symmetry breaking, $M_{Z, W }\sim
v_H$ and $v_H \simeq 10^2~GeV$.  A fine tuning of coupling constants
is necessary to obtain $v_H << v_{\Phi}$. This is the gauge hierarchy
problem which we explained above.
\b
\no{\it 7.2. ~Predictions of the $SU(5)$ model}
\s
The electromagnetic charge, $Q$, is a generator of $SU(5)$.  As such $Tr Q
= 0$, i.e. the sum of the charges
$Q$ over any representation must be zero.
 For example, applying this
 requirement to the $\bar{5}$ in Eq. (118), one obtains
$$ 3Q_{d^c} + Q_e = 0~. \eqno(122)$$
Thus $Q_e = -1$ implies $Q_{d^c} = 1/3$, hence $Q_d = - 1/3$,
 and since $Q_u = Q_d + 1 =
2/3$, we obtain charge quantization, i. e., $Q_p = 2Q_u + Q_d = 1$.
\b
Let us call $I_3$ the $5 \times 5$ matrix in Eq. (116)
that contains $\tau^3/2 = T_3$,
the third generator of $SU_L(2)$. Then $Q = T_3 + Y$ translates in
$SU(5)$ into
$$ Q = I_3 + c I_0~, \eqno(123)$$
where $c$ is a constant.  Thus $Y = cI_0$.
The difference between $Y$, the standard
$U_Y(1)$ generator, and $I_0$, the
corresponding $SU(5)$ generator, is due to the different normalization
 of both generators:  $I_0$ must have
 the same normalization as $I_3$,
$$ Tr (I_3^2) = Tr~ (I_0^2)~=~ r ~,\eqno(124)$$
according to Eq. (28), while the normalization of $Y$ and $T_3$ are
independent.  We can compute, the constant
$c$ using the generators in any
representation, for example the $\bar{5}$ in Eq. (118).  The
generator $Y$ applied to it, is
$$ Y_{{\bar{5}}}~~=~~ \pmatrix{ 1/3 & 0 & 0 & 0 & 0 \cr
0 & 1/3 & 0 & 0 & 0 \cr 0 & 0 & 1/3 & 0 & 0 \cr
0 & 0 & 0 & -1/2 & 0 \cr
0 & 0 & 0 & 0 & -1/2 \cr}~. \eqno(125)$$
Plugging $Y_{\bar{5}}$ (Eq. (125)) and $I_0$ (Eq. (116)), in the relation
$Y = c I_0$, we obtain $1/3 = cA = c ( - 2/\sqrt{60})$, thus
$$ c = - \sqrt{{3\over 5}}~. \eqno(126)$$
In each covariant derivative, the corresponding
coupling constant multiplies each gauge boson and generator.
 The terms containing the bosons $W^3_{\mu}$ and $B_{\mu}$ in
the covariant derivatives of
$SU(5)$ and $SU(2) \times U(1)$ (Eq. 34) must coincide at the
unification scale $M_X$:
$$ g(M_X)~~I_3 W^3_{\mu} + g (M_X)~~ I_0 B_{\mu} =
 g_2 (M_X)~~ I_3 W^3_{\mu}
+ g_1 (M_X)~~ YB_{\mu}~~. \eqno(127)$$
Notice that the $SU(5)$ expression has a unique coupling $g (M_X)$.
 From Eq. (127) we get $g (M_X) = g_2 (M_X)$ and, using $Y = cI_0$,we get $g
(M_X) c = g_1 (M_X)$. Consequently, using the value of
$c$ in Eq. (126), we obtain
$$ {g^2_1 (M_X)\over g^2_2 (M_X)} = {3\over 5}~. \eqno(128)$$
Since $\sin^2 \theta_W \equiv g^2_1/(g^2_2 + g^2_1)$, at the
unification scale $M_X$ we have
$$ \sin^2 \theta_W \bigg\vert_{M_{X}} = {c^2\over 1 + c^2} = {3\over
8}~. \eqno(129)$$
We can convince ourselves that this prediction is common to all GUT's
for which the fermions only include the standard fermions of one
generation (plus possibly singlets of the S.M. group).  This is because
  the equality $Tr (I_0^2) = Tr (I^2_3)$ holds in any GUT
  ($Tr$ indicates the sum
over all fermions in one family). Thus, summing the hypercharges  and the
third isospin components of all the fermions in one standard family, i.e.
 $\sum_f Y_f^2$ and $\sum_f I_3^2$, we can compute the ratio
$${Tr (Y^2)\over Tr (I^2_0)} = {Tr (Y^2)\over Tr (I_3^2)} = {\sum_f
Y_f^2\over \sum_f I_3^2} = {5\over 3}~. \eqno (130)$$
Since at the scale $M_X$ we have $g_1^2 Tr (Y^2) = g^2_2 Tr (I_0^2)$ (Eq.
 (127)), we obtain
again Eq. (128) and thus the prediction in Eq. (129).

In order to compare the prediction in Eq. (129) with the measured value of
$\sin^2 \theta_W$ one has to renormalize the coupling constants from
$M_X$ to $M_Z$.  One
obtains $^{59}$,
$$ \sin^2 \theta_W \bigg\vert_{M_{Z}}
 \simeq  0.207-0.213~,~~~ M_X \simeq (2 - 7)~ 10^{14}~GeV~, \eqno(131)$$
both too low to be acceptable.  The present value of $ \sin^2 \theta_W$
is $\simeq 0.23$, and the actual calculation of the proton
life time gives $^{4}$
$$ \tau_{p \to e^+ \pi^0}~\simeq~ 10^{31 \pm 0.7} y~~\left({M_x\over 4.6
\times 10^{14}~GeV}\right)^4~, \eqno(132)$$
which combined with the experimental bound,
$ \tau_{p \to e^+ \pi^0} > 10^{33} y$, $^{58}$ yields $M_X >$ $1.5$
$10^{15}$ $GeV$.
\b
The fermion masses come from the Yukawa couplings of
$f_{\bar{5}}$ and $f_{10}$ (Eq. (118)) with the Higgs field $H$.
 For each generation, these couplings are,
$$ {\cal{L}}_Y = \lambda_1 {\bar{f}}_{{\bar{5}} {\alpha}} f_{10 {\alpha}
{\beta}} H_{\beta} + \lambda_2 \epsilon_{\alpha \beta \gamma \delta
\epsilon} f_{10{\alpha} {\beta}} f_{10{\delta} {\gamma}} H_{\epsilon}
\eqno(133)$$
where $\alpha, \beta, \gamma, \delta, \epsilon = 1,2,3,4,5.$  Replacing
$H~ {\rm by}~ < H >$  (Eq. (121)), one gets
$$ {\cal{L}}_{\rm mass} ~\simeq~ \lambda_1 v_H
({\bar{d}} d + \bar{e} e)~+~ \lambda_2 v_H ({\bar{u}} u)~~ \eqno(134)$$
for each generation.  Thus, at the scale $M_X$, one predicts
$$ m_d = m_e~,~~m_s = m_{\mu}~,~~ m_b = m_{\tau} ~.\eqno(135)$$
The renormalization of the Yukawa couplings change these ratios.  The
prediction $^{60,~4}$
$$ {m_b\over m_{\tau}}~=~ 2.8~ -~ 2.9~,\eqno(136)$$
is considered a success, but the  ratios $m_s/m_{\mu}$ and $m_d/m_e$ fail.

Because
of the bounds on $p-$decay, $\sin^2 \theta_W$  and fermionic mass
ratios, the minimal $SU(5)$ model presented here is considered to be
ruled out.  So we need to go to more complicated models with larger
symmetries and/or more fields.  One of the ways of enlarging the symmetry
is to include supersymmetry.
\b
\no {\it 7.3. ~Supersymmetric GUT's}
\s
Precision measurements at LEP have shown that the standard model
coupling constants, when extrapolated to higher energies without
introducing any new physics, do not meet at one point$^{59,~61}$.  This is
additional evidence of the failure of simple GUT's as the minimal
$SU(5)$ model.  The same calculation in the M.S.S.M.  (instead of the S.M.)
shows that the three couplings do meet at one point, at a unification
scale $M_X \simeq O(10^{16} GeV)$, large enough to suppress proton decay
mediated by the exchange of lepto-quark gauge bosons$^{59,~61}$.
 Notice that the
extrapolation of coupling constants from $M_Z$ to $M_X$ is independent
of the GU model, since all fields with mass $\sim M_X$ drop out of the
renormalization group equation.  This extrapolation depends only on the
low energy spectrum, and the simplest viable low energy supersymmetric
model is the M.S.S.M.

Also the prediction for $\sin^2 \theta_W$ is
better.  Recall that starting from 3/8 at the unification scale, the
renormalized value of $\sin^2 \theta_W$ at the $M_Z$ scale is too low,
when the low energy spectrum is that of the S.M.  With the M.S.S.M.
spectrum instead, the prediction is compatible with the experimental
data$^{59,~ 61}$.  These arguments by no means
``prove" supersymmetry,  since
similar results can be obtained in non-supersymmetric GUT's, as we will
see below.  However, they support the idea of supersymmetry.

 There are some  potentially serious problems
 in supersymmetric GUT's. The renormalization
 of the $ {m_b/ m_{\tau}}$ ratio in minimal models,
  starting from 1 at the
GUT scale, leads to the very good
prediction $^{60}$ $ {m_b}= 4.9 \pm 0.1~GeV$, but again,
other fermion mass  ratios fail.
Another potentially serious problem is  proton
decay mediated by ``dimension 5 operators"$^{62}$. While
the high value of
$M_X$ yields a safe value for the $p$-lifetime due to the exchange of
lepto-quarks, the exchange of squarks and gauginos/Higgsinos in a box
diagram is dangerous.  The $B$-violating operator,
$$ {\cal{L}}_{d = 5} = {\tilde{q}~ \tilde{q}~ q~ \ell\over M_X}~,
\eqno(137)$$
couple two bosons, two squarks, with two fermions,
 a quark and a lepton, through
the exchange of a fermion, a heavy coloured Higgsino
 of mass $M_X$.  The mass
dimension of these operators is $d = 5$ (1 for each
boson and 3/2 for each
fermion).  The box diagram that mediates $p$-decay
 through $q q \to \bar{q} \bar{\ell}$,
consists of $q q \to \tilde{q} \tilde{q}$ (through
the exchange of a light gaugino) and $\tilde{q}
\tilde{q} \to \bar{q}
\bar{\ell}$ $\bigl($through the coupling in
Eq. (137)$\bigr)$.
The exchange of lepto-quark
gauge bosons induces $p$-decay with a lifetime
 $\tau_p \sim M_X^{-4}$, while the exchange of a
coloured Higgsino (a fermion) yields a much shorter
lifetime $\tau_p \sim M_X^{-2} << M_X^{-4}$.  Thus, one predicts
$$ \tau_{p \to \bar{\nu} K^+} \simeq 10^{29 \pm 4} y ~,\eqno (138)$$
quite low compared with the experimental lower bound of $10^{33}y$.
Therefore, this decay mode provides strong constraints on many
supersymmetric models, and excludes some of them$^{63}$.

 There are already
models in which the $B$-violating $d = 5$ operators,
Eq. (113), are absent.  This is the case of the
``flipped $SU(5) \times U(1)$" model$^{64}$.
 This is not truly a G.U.T., since the gauge
group  $SU(5) \times U(1)$ is not a simple group.  The matter fields
in this model are:

$$\eqalign{ &(\bar{5}, - 3/2)
:\left(\matrix{u^c \cr u^c \cr u^c \cr \nu \cr e^-\cr}
\right)_L\!\!\!; ~~~~~(10, 1/2):\pmatrix{0 & d^c  & -d^c & d &
u \cr -d^c & 0  & d^c & d & u \cr d^c_y &
-d^c & 0 & d & u \cr -d & -d & -d & 0 & \nu^c \cr -u & -u & -u &
-\nu^c & 0 \cr}_L\!\!\!; \cr
& \cr
 &(1,5/2):e^c;~~~~~~~~~~~~~~~(1,0):N~. \cr}\eqno(139)$$
Here the colour indices have been neglected, the first number in the
parenthesis indicates the representation of $SU(5)$ and the second the
$U(1)$ charge.  Notice that the assignments of $u$ and $d$ and those of
$\nu$ and $e$, have been ``flipped" with respects to the assignments in
$SU(5)$ GUT, Eq. (118).  One needs an extra lepton, $N$, and four Higgs
multiples, $\Phi, \bar{\Phi}, H$ and $\bar{H}$,
in the representations $(10,
1/2), (\bar{10}, -1/2), (5,-1)$ and $(\bar{5}, 1)$ respectively, to break
spontaneously the symmetry.  In this model, the $d = 5$ operators in Eq.
(137) depend on the existence of a mixing between the
coloured components of $H$ and $\bar{H}$ (that we call
$ {H_c}$ and ${\bar{H}}_c$). Actually, they depend on the mixing of their
supersymmetric partners, the Higgsinos ${\tilde{H}}_c $ and
$\buildrel{\simeq}\over {H_c}$. In this model,
  there is a
mechanism that insures that the coloured Higgs fields in $H$ and $\bar{H}$ are
heavy, while the non coloured ones remain light (and constitute the
standard doublet Higgs field). It happens that this mechanism
also suppresses the ${\tilde{H}}_c -
\buildrel{\simeq}\over {H_c}$  Higgsino mixing,
and thereby avoids the danger of $p$-decay through $d = 5$ operators.

As we mentioned above, the confluence of the three coupling constants at
one point is not proof of supersymmetry, since it can be also achieved in
non-supersymmetric GUT's with intermediate breaking scales.  Let us
mention $SO(10)$ models with two breaking scales: $M_U$, the usual
unification scale, and $M_I$, another intermediate scale $^{59,~65}$.
  At $M_U$ the
$SO(10)$ group breaks into a left-right symmetric group, either
$$SU_L (2) \times SU_R (2) \times SU(4)~, \eqno(140)$$
or
$$SU_L(2)\times SU_R(2) \times U_{B-L}(1) \times SU_C(3)~, \eqno(141)$$
that break into the S.M. at the scale $M_I$.  The correct value of
$\sin^2 \theta_W$ and a safe $p$-lifetime can be obtained with $M_U
\simeq 10^{16} GeV,~ M_I \simeq 10^{12}~GeV$ for the first case, Eq.
(140), or $M_U \simeq 10^{16}~GeV,~ M_I \simeq 10^{10}~GeV$ for the
second case, Eq. (141).

$SU(10)$ models $^{66,~4}$ fit the  15 Weyl spinors of
 a whole standard fermion family, into a
representation of dimension 16.  Thus, an additional particle, a
right-handed neutrino, must be added.  These models, therefore, have
naturally a ``see-saw"$^{23}$ mechanism to produce neutrino masses.  This
mechanism is based on the fact that a mass matrix of the form
$$ ({\overline{{\nu}_L}} ~~~ {\overline{{{\nu}^c}_L}})
\pmatrix{ 0 & m_D \cr m_D & M
\cr} \pmatrix{{\nu^c}_R \cr {\nu}_R \cr} ~, \eqno(142)$$
with $M >> m_D$, has one light eigenvector, mainly $\nu_L$, with mass
$m_D^2/M$, and a heavy eigenvector, mainly $\nu_R$, with mass $M$.  $M$
is a $\nu_R$ Majorana mass,$~~M~ {\overline{{\nu}^c_L}} \nu_R$, and $m_D$ is a
Dirac mass, $~m_D ~{\overline{{\nu}_L }}\nu_R =
m_D ~\overline{{{\nu}^c}_L} {{\nu}^c}_R $.
With three generations, each fermion in Eq. (142) becomes a vector of
three components, $~m_D$ and $M$ become $ 3 \times 3$ matrices,
and the $ 6 \times 6$ mass matrix has six eigenvectors,
 (usually) three light and three heavy,
$$ {m_{D1}^2\over M_1},~~ {m_{D2}^2\over M_2},~~ {m_{D3}^2\over M_3},~~
M_1, M_2, M_3~. \eqno(143)$$
There are many different versions of this mechanism.  For example, the
Dirac masses $m_{Di}$ could be similar to the up-quark masses, or to the
charged lepton masses (or even something different).  The neutrino mass
hierarchy depends on this choice since $m_t/m_c \simeq 100$ while
$m_{\tau}/m_{\mu} \simeq 17$.  In a ``quadratic see-saw", the three heavy
masses $M_i$ are similar, i. e. $M_i \simeq M$ for $i = 1,2,3$. Consequently,
the hierarchy of light neutrino masses is that of $m^2_{Di}$, i.e.
$m_{\nu_{1}} : m_{\nu_{2}} : m_{\nu_{3}} \simeq m^2_u : m_c^2 : m^2_t$
(or $m_e^2 : m_\mu^2 : m_{\tau}^2)$.  In a ``linear see-saw", the
hierarchy of the heavy masses  $M_i$
coincides with that of $m_{Di}$, i.e.
$M_1 : M_2 : M_3 \simeq m_u : m_c : m_t$.
As a consequence, also the ratio of light
neutrino masses is linear in $m_D$.

 $SO(10)$ models with intermediate scales
$M_I$, naturally have the heavy Majorana masses
at $M_I$,  $M_i \simeq M_I \simeq 10^{10-12}~ GeV$.
As a result they easily accommodate the MSW $^{67}$
 solution to the solar neutrino problem, that requires
 $\vert m^2_{\nu_{i}} -
m^2_{\nu_{e}} \vert^2 \simeq 10^{-6}~eV$ $^{21}$,
by yielding $m_{\nu_{e}} < <
m_{\nu_{\mu}} \simeq 10^{-3} eV$. In this case,
the value of $m_{\nu_{\tau}}$,  that could
also have important implications, depends on the type of see-saw.
In quadratic see-saw models one may obtain $m_{\nu_{\tau}} \simeq 10~
eV$, so that $\nu_{\tau}$ could be a good D.M. candidate.  In linear
see-saw models one may obtain $m_{\nu_{\tau}}
\simeq ~10^{-1} eV$, a good
value for solving the ``atmospheric neutrino problem"
 $^{22}$, (a deficit of
$\nu_{\mu}$ with respect to $\nu_e$ in cosmic rays) through $\nu_{\mu} -
\nu_{\tau}$  oscillations .
\b
\no {\bf 8. ~Brief Comment on Composite Models $^{68}$}
\b
The basic idea is that a few fundamental particles may form bound states
only in allowed configurations, i.e., the combinations of colour,
charge,
etc. of the S.M. particles.  In particular, excitations of the
fermionic bound states would explain the existence of more than one
family.  There is an obvious problem in considering quarks and leptons
as bound states, namely
their masses are much smaller than their inverse radii. This is contrary
to all bound states we know.
For
example, $(g-2)$ measurements give an upper bound of $10^{-19} cm$ to
the size of electrons, thus:
$$ m_e \simeq 0.5 MeV < < 10^{19}cm^{-1} \simeq 10^3 TeV \simeq
\Lambda_H~. \eqno(144)$$
Here $\Lambda_H$ is the scale of the confining force ``hypercolour",
that binds the components, the ``preons".  Eq. (144) shows that
the origin of the radii (the confinement scale
$\Lambda_H$) and the origin of masses $m$
must be different. When bound states form, at the energy scale
$\Lambda_H$, some of them must remain massless due to a symmetry
(usually chiral symmetry)$^{69}$ that is
spontaneously broken at a lower scale
$m$, at which masses appear, $m < < \Lambda_H$ .  If a chiral symmetry is
present among bound states, it must be also present among preons that,
therefore, must be massless. Consequently,
we have massless fermions forming
massless composites.  't Hooft $^{69}$ has formulated
an important constraint on
composite models often called 't Hooft's ``anomaly
 matching condition".
It states that for chiral symmetry to be unbroken
in the formation of
composites (so producing massless bound states),
it is necessary that the
anomalies of axial currents must be the same when computed using
both sets of
massless fermions, i.e., preons and bound states.  This is not an easy
condition to fulfill, since preons and bound states usually are
different in number and belong to different representations.  Realistic
models $^{70}$ are very complicated and lack  predictivity.
\b
Composite gauge bosons should reproduce with extreme accuracy the
relations between different couplings found in gauge theories.  This is
very difficult. The cancellations of different diagrams that
insures unitarity in processes such as $e^+ e^- \to W^+ W^-$, give good
bounds on possible departures from a gauge structure.
Moreover, if $W$ and $Z$ are composites one expects to
find also excited states, $W^{\prime}, W^{\prime\prime},\cdots $ and
$Z^{\prime}, Z^{\prime\prime}, \cdots$. We would
need a mechanism to insure that only the lightest bound states have
masses $~M_{W, Z} << \Lambda_H$, otherwise the values of $\sin^2
\theta_W$ and the $\rho$-parameters should depart from the S.M.  As the
experimental agreement with the S.M. improves, there is progressively
less room for composite $W$ and $Z$ bosons ${68}$.
\b
\no {\bf 9. ~Acknowledgements}
\b
I thank the organizers of the Lake Louise Winter Institute
for the opportunity
to participate in such a pleasant school.  This work was supported in
part by the DOE under grant DE-FG03-91ER 40662 Task C.
\b
\no{\bf 10. ~References}
\b
\item{1.} S. L. Glashow, Nucl. Phys. 22 (1961) 579; S.
Weinberg, Phys. Rev. Lett. 19, (1967) 1264; A. Salam,
{\it Proceedings of the 8th. Nobel Symposium}, Ed. N. Swartholm (Stockholm,
1968).
\b
\item{2.} D. Gross and F. Wilczek, Phys. Rev. Lett. 30 (1973) 1343; S.
Weinberg, Phys. Rev. Lett. 31 (1973), 494; H. Fritzsch, M. Gell-Mann and
H. Leutweyler, Phys. Lett. 47B (1973) 365.
\b
\item{3.} J. Bjorken and S. Drell, {\it Relativistic Quantum
Mechanics} (McGraw-Hill, 1964).
\b
\item{4.} For  reviews see P. Langacker, Phys. Rep. 72 (1981) 185; {\it
Nith Worshop on Grand Unification} ed. R. Barloutaud
 (Worl Scientific, Singapore, 1988) p. 3; G.G. Ross,
  {\it Grand Unified Theories} (Benjamin, New York, 1984);
   R. N. Mohapatra, {\it Unification and Supersymmetry}
   (Springer, New York, 1986,1992).
\b
\item{5.} R. Peccei and H. Quinn, Phys. Rev. Lett. 38 (1977) 1440 and
Phys. Rev. D 16 (1977) 1791; F. Wilczek, Phys. Rev. Lett. 40 (1978) 279;
S. Weinberg, Phys. Rev. Lett. 40 (1978) 223.
\b
\item{6.} M. Gell-Mann, California Inst. of Technology Syncrotron Lab.
Report N$^o~$  CTSL -20 (1961) (reprinted in M. Gell-Mann and Y. Ne'eman,
{\it The Eightfold Way} (Benjamin, New York, 1964); Y. Ne'eman, Nucl.
Phys. 26 (1961) 222.
\b
\item{7.} Y. Nambu, Phys. Rev. Lett. 4 (1960) 380; J. Goldstone, Nuovo
Cimento 19 (1961) 154.
\b
\item{8.} P. Higgs, Phys. Rev. Lett. 12 (1964) 132 and Phys. Rev. l45
(1966) 1156; F. Englert and R. Brout, Phys. Rev. Lett. 13 (1964) 321; G.
Guralnik, G. Hagen and T. Kibble, Phys. Rev. Lett. 13 (1964) 585.
\b
\item{9.} M. Kobayashi and M. Maskawa, Prog. Theor. Phys. 49 (1973)
652.
\b
\item{10.} L. Susskind, Phys. Rev. D 20
(1979) 2619; S. Weinberg,  Phys. Rev. D 16 (1976) 974,
Phys. Rev. D 19 (1979) 1277; for  reviews see, for example
E. Fahri and L. Susskind, Phys. Rep. C 74 (1981) 227;
 S. King, {\it Beyond the Standard Model}
(Carleton Univ., Ottawa, June 1992), CERN-preprint-TH. 6617/92)
\b
\item{11.} D. Volkov and V.P. Akulov, Phys. Lett. 46B (1973) 49; J. Wess and
B. Zumino, Nucl. Phys. B 70 (1974) 39.
See for example, P. Fayet and S. Ferrara, Phys. Rep. C32
(1977); J. Wess and J. Bagger, {\it  Supersymmetry and Supergravity}
(Princeton U. Press, 1983).
\b
\item{12.} H. Georgi and S. Glashow, Phys. Rev. Lett. 32 (1974) 438.
\b
\item{13.} S. Dimopoulos and H. Georgi, Nucl. Phys. B193 (1981) 150; N.
Sakai, Zeit. fur Phys. C11 (1982) 153.
\b
\item{14.} F. Wilczek and A. Zee, Phys. Rev. Lett. 42 (1979) 421.
\b
\item{15.} See for example J. Bagger, S. Dimopoulos and E. Masso, {\it Int.
Conf. on High Energy Physics} (Leipzig, Germany, 1984) Vol. 1, 12.
\b
\item{16.} E. Cremmer at al., Phys. Lett. 79B (1978)
 23 and Nucl. Phys.
B147 (1979) 105; See for example P. Van Nieuwenhuizen,
Phys. Rep. 32C (1977)
189; P. Nilles, Phys. Rep. 110C (1984) 2; Wess and Bagger in Ref.11.
\b
\item{17.} See, for example, E. Kolb and M. Turner, {\it The Early
Universe} (Addison-Wesley, Reading, MA. 1990) and references therein.
\b
\item{18.} See, for example, D. Bennett et al.
(MACHO collaboration),{\it
16th. Texas Symposium, TEXAS-PASCOS 92} (Berkeley,1992)
\b
\item{19.} K. Olive, D. Schramm, G. Steigman and T. Walker, Phys. Lett. B236
(1990) 454.
\b
\item{20.} J. Ellis, D. Nanopoulos, L. Roszkowski and D. N. Schramm,
Phys. Lett. B245 (1990); L. Krauss, YCTP-P4-91 (1991) and YCTP-P36-91 (1991);
L. Roszkowski, Phys. Lett. B278 (1992) 147.
\b
\item{21.} See, for example, S.A. Bludman, N.Hata  D.C. Kennedy and
P.G. Langacker, Phys. Rev. D47 (1993) 2220.
\b
\item{22.} K.S. Hirata et al., Phys. Lett. B280 (1992) 146;
 R. Becker-Szendy
et al., Phys. Rev. D46 (1992) 3720.
\b
\item{23.} M. Gell-Mann, P. Ramond and R. Slansky, {\it Supergravity}
(ed. F. Van Nieuwenhuizen and D. Freedman
(North Holland, Amsterdam, 1979) p. 315;
T. \break Yanagida {\it  Proc. of the Workshop
 on Unified Theory and Baryon Decay in the Universe}
 (KEK, Japan, 1979), Prog. Theo. Phys. B135 (1978) 66.
\b
\item{24.} S. Dimopoulos and L. Susskind,
 Nucl. Phys. B 155 (1979) 237; E. Eichtein and K. Lane,
  Phys. Lett. 90 B (1980) 237.
\b
\item{25.} B. Holdom, Phys. Rev. D24 (1981) 1441;
Phys. Rev. Lett. B150 (1985.
301.
\b
\item{26.} W. Bardeen, C. Leung and S. Love,
 Phys. Rev. Lett. 56 (1986) 1230;
T. Appelquist, D. Karabali and L. Wijewardhana,
 Phys. Rev. Lett. 57 (1986) 957;
T. Appelquist and L. Wijewardhana, Phys. Rev. D 35 (1987) 774
and Phys. Rev. D
36 (1987) 568.
\b
\item{27.} For a review see, for example, T Appelquist,
YCTP-P23-91, published in {\it Proc. of Particles, Strings
and Cosmology} (Boston 1990) p.315.
\b
\item{28.} T. Appelquist, T. Takeuchi, M. Einhorn and
L. Wijewardhana, Phys. Lett. B 220 (1989) 223; B. Holdom,
Phys. Lett. B 236 (1990) 327.
\b
\item{29.} M. Peskin  and T. Takeuchi, Phys. Rev. Lett. 65
(1990) 964 and
Phys. Rev. D46 (1992) 381; J. Ellis, G. Fogli and E. Lisi,
 Phys. Lett. B 274
(1992) 454.
\b
\item{30.} T. Appelquist and G. Triantaphyllou,
 Phys. Lett. B 278 (1992) 345; R. Sundrum and S. Hsu,
  Nucl. Phys. B 391 (1993) 127.
\b
\item{31.} T. Appelquist and J. Terning,
Yale U. preprint YCTP-P7-93, April 1993.
\b
\item{32.} R. Chivukula, M. Dugan and M. Golden,
 Phys. Lett. B 292 (1992) 435 and Phys. Rev. D47
 (1993) 2930; T. Appelquist and J. Terning, Phys. Rev. D47
 (1993) 3075.
\b
\item{33.} Y. Nambu, {\it New Theories in Physics},
Proc. XI Warsaw Symposium on Elementary Particle Physics
(ed. Z. Ajduk et al., World Scientific, 1989);
V. Miransky, M Tanabashi  and Y. Yamawaki, Phys. Lett. B 221
(1989) 177 and Mod. Phys. Lett. A4  (1989) 1043.
\b
\item{34.} W. Bardeen, C. Hill and M. Lindner, Phys. Rev. D41 (1990) 1647.
\b
\item{35.} M. Suzuki, Mod. Phys. Lett. A5 (1990) 1205;
S. King and S. Mannan, J. Mod. Phys. A6 (1991) 4949;
 A. Hasenfratz et al., Nucl. Phys. B365 (1991) 79.
\b
\item{36.} G. Farrar and P. Fayet, Phys. Lett. 76B (1978) 575;
 G. Farrar and
S. Weinberg, Phys. Rev. D27 (1983) 2732.
\b
\item{37.} L. Hall and M. Suzuki, Nucl. Phys. B 231 (1984) 419;
 R. Barbieri
et al., Phys. Lett. B 238 (1990) 86.
\b
\item{38.} C. Aulak and R.N. Mohapatra, Phys. Lett. 119b (1983) 136;
 G. G. Ross and J. Valle, Phys. lett. 151B (1985) 375;
 J. Ellis et al., Phys. Lett. 150B (1985) 142; J. Valle,
 Phys. Lett. B 196 (1987) 157; M. C.
Gonzalez-Garcia and J. Valle, Nucl. Phys. B 355 (1991) 330;
A. Masiero and J. Valle, Phys. Lett. B 251 (1990) 273.
\b
\item{39.} For a review see, for example, J. Valle,
 {\it Neutrino 92} (Granada, Spain, June 1992),
  CERN preprint TH.6626/92.
\b
\item{40.} H. Haber and G. Kane, Phys. Rep. 117 (1985) 75;
J. Gunion and H. Haber, Nucl. Phys. B 272 (1986) 1;
R. Barbieri, Riv. Nuovo Cim. 11 (1988)
No. 4.
\b
\item{41.}  For a recent review  see, for example,
 F. Zwirner, {\it Texas/PASCOS Conference} (Berkely, Dec. 1992)
  and CERN-TH.6357/91.
\b
\item{42.} J. Ellis and F. Zwirner, Nucl. Phys. B 338 (1990) 317.
\b
\item{43.} K. Griest and L. Roszkowski, Phys. Rev. D46 (1992) 3309.
\b
\item{44.} J. Ellis, J. Gunion, H. Haber, L. Roszkowski and F. Zwirner,
Phys. Rev. D39 (1989) 844; R. Flores, K. Olive and D. Thomas, Phys. Lett.
 B 245 (1990) 509 and Phys. Lett. B 263 (1991) 263.
\b
\item{45.} G. Gelmini, P. Gondolo and E. Roulet, Nucl. Phys. B 351 (1991)
623.
\b
\item{46.} I. Adachi et al. (TOPAZ coll.), Phys .Lett. B 218 (1989)
 105 and KEK- preprint -89-037 (1989).
\b
\item{47.} D. Decamp et al (ALEPH coll.), Phys. Rep. 216 (1992) 253.
\b
\item{48.} G. Ridolfi, G. Ross and F. Zwirner,
{\it Proc. of the LHC Workshop}
(Aachen, Germany, G. Jarlskog and D. Rein eds.,
 Oct. 1992) Vol. II, p. 605.
\b
\item{49.}  J. Ellis, G. Fogli and E. Lisi,
 Nucl. Phys. B 393 (1993) 3.
\b
\item{50.}  H. Baer, X. Tata and J. Woodside,
Phys. Rev. D 44 (1991) 207.
\b
\item{51.}  F. Abe et al. (CDF coll.),
 Phys. Rev. Lett. 69 (1992) 3439.
\b
\item{52.} A. Bottino et al., Phys. Lett. B 265 (1991) 57; A. Bottino et al.
Mod. Phys. Lett. A 7 (1992) 733.
\b
\item{53.} J. Ellis and L. Roszkowski, Phys. Lett. B 283 (1992) 252.
\b
\item{54.} K. Olive and M. Srednicki, Phys. Lett. B 230 (1989) 78;
 Nucl. Phys. B 555 (1991) 208; K. Griest, M. Kamionkowski and
  M. Turner, Phys. Rev. D 41 (1990) 3565; M. Kamonkowski,
   Phys. Rev. D 44 (1991) 3021.
\b
\item{55.}  Y. Okada, M. Yamaguchi and T. Yanagida,
 Prog. Theor. Phys. Lett. 85  (1991) 1; J. Ellis, G. Ridolfi
 and F. Zwirner, Phys. Lett. B 257 (1991) 83; H.
Haber and R. Hempfling, Phys. Rev. Lett. 66 (1991) 1815.
\b
\item{56.} Z. Kunszt and F. Zwirner, Nucl. Phys. B 285 (1992) 3.
\b
\item{57.} J.R. Espinosa and M. Quiros, Phys. Lett. B 266 (1991) 389,
 Phys.
Lett. B 279 (1992) 92 and Phys. Lett. B 302 (1993) 51.
\b
\item{58.} For a review see R. Barloutaud,
 {\it TAUP' 91} (Toledo, Spain, Sept. 1991),
 Nucl. Phys. B (Proc. Suppl.) 28A (1992) 437.
\b
\item{59.} P. Langacker and M. Luo, Phys. Rev. D44 (1991) 817;
P. Langacker,
UPR-0539-T, May 1992.
\b
\item{60.} D. Nanopoulos and D. Ross, Phys. Lett. 108 B (1982)
 351 and Phys. Lett. 108 B (1982) 99.
\b
\item{61.} J. Ellis, S. Kelley and D.V. Nanopoulos,
Phys. Lett. B 249 (1990) 441; U. Amaldi, W. de Boer
and H. Furstenau, Phys. Lett. B 260 (1991) 447;
F. Anselmo, L. Cifarelli, A. Peterman and A. Zichichi,
 Nuovo Cim. 104A (1991)
1817; P. Langacker and N. Polonsky, Phys. Rev. D47 (1993) 4028.
\b
\item{62.} S. Weinberg, Phys. Rev. D 26 (1982) 287;
 N. Sakai and T. Yanagida, Nucl. Phys. B 197 (1982) 533;
 S. Dimopoulos, S. Raby and F. Wilczek, Phys. Lett. 112B (1982) 133;
  J. Ellis, D.V. Nanopoulos and S. Rudaz, Nucl. Phys. B 202 (1982) 43.
\b
\item{63.} See, for example, R. Arnowitt and P. Nath,
 {\it Lake Louise Winter Inst.} (Lake Louise, Canada,
  Feb. 1992), CTP-TAMU-39/92 and references therein;
  also Phys. Lett. B 299 (1993) 58 and CTP-TAMU-23-93.
\b
\item{64.} S. Barr, Phys. Lett. B 112 (1982) 219,
 Phys. Rev. D 40 (1989) 2457;
J. Derendinger J. Kim and D. V. Nanopoulos, Phys. Lett.
B 139 (1984) 170;
I. Antoniadis, J. Ellis, J. Hagelin and D. V. Nanopoulos,
Phys. Lett. B 194 (1987), Phys. Lett. B 231 (1989) 65; J. Ellis,
 J. Hagelin, S. Kelley and D. V.
nanopoulos, Nucl. Phys. B 311 (1988) 1.
\b
\item{65.} R.N. Mohapatra and  M.K. Parida,
UMD-PP-92-170 and Phys. Rev. D
47 (1992) 264.
\b
\item{66.} H. Georgi, {\it APS Particles and Fields}
(ed. C.E. Carlson,
AIP, New York, 1975) p.575; H. Fritzsch and P. Minkowski,
 Ann. Phys. 93 (1975) 193.
\b
\item{67.} L. Wolfenstein, Phys. Rev. D 25 (1978) 2369;
 S. P. Mikheyev and
A. Yu. Smirnov, Yad. Fiz. 42 (1985) 1441
(Sov. J. Nucl. Phys. 42 (1985) 913).
\b
\item{68.} For reviews see, for example, M. Peskin,
 {\it Proc. Int. Symposium on Lepton and Photon
 Interactions at Haigh Energy} (ed. M. Koruma and K.
Takahashi, 1985);  R. D. Peccei, {\it Proc. of the
 Lake louise Winter Institute}, (Lake Louise, Canada, 1987)
  p. 564, DESY-87-050 (1987).
 G. t' Hooft in {\it Recent Developments in Gauge
 Theories}, Cargese Lectures 1979, (eds. G. 't Hooft et al.,
  Plenum, New York, 1980)
\b
\item{70.} For a recent attempt see, for example, S. Yu. Khlebnikov
and R. D. Peccei, UCLA-92-TEP-49 preprint (1992)

\end
\bye